\documentclass[11pt,a4paper]{article}
\usepackage{jcappub}                    
\usepackage{amsmath,amssymb,mathtools}  
\usepackage{soul}                       
\usepackage[usenames,dvipsnames]{xcolor}

\newcommand{\dd}{\text{d}}

\setstcolor{red}    
\setulcolor{red}    
\allowdisplaybreaks 

\title{Logarithmic divergences in the $k$-inflationary power spectra computed through the uniform approximation}

\author{Allan L. Alinea$^{a)}$,} 
\author{Takahiro Kubota$^{a), b), c)}$,}
\author{and Wade Naylor$^{a), d)}$}

\affiliation{ $a)$ Department of Physics, Osaka University, Toyonaka, Osaka 560-0043, Japan}
\affiliation{ $b)$ CELAS, Osaka University, Toyonaka, Osaka 560-0043, Japan}
\affiliation{ $c)$ Kavli IPMU (WPI), The University of Tokyo, 5-1-5 Kashiwa-no-Ha, Kashiwa City, Chiba 277-8583, Japan}
\affiliation{ $d)$ International College, Interdisciplinary Research Building, Osaka University, Toyonaka, Osaka 560-0043, Japan}

\emailAdd{alinea@het.phys.sci.osaka-u.ac.jp}
\emailAdd{kubota@celas.osaka-u.ac.jp}
\emailAdd{naylor@phys.sci.osaka-u.ac.jp}
 
\abstract{We investigate a calculation method for solving the Mukhanov-Sasaki equation in slow-roll $k$-inflation based on the uniform approximation (UA) in conjunction with an expansion scheme for slow-roll parameters with respect to the number of $e$-folds about the so-called \textit{turning point}. Earlier works on this method has so far gained some promising results derived from the approximating expressions for the power spectra among others, up to second order with respect to the Hubble and sound flow parameters, when compared to other semi-analytical approaches (e.g., Green's function and WKB methods).  However, a closer inspection is suggestive that there is a problem when higher-order parts of the power spectra are considered; residual logarithmic divergences may come out that can render the prediction physically inconsistent. Looking at this possibility, we map out up to what order with respect to the mentioned parameters several physical quantities can be calculated before hitting a logarithmically divergent result. It turns out that the power spectra are limited up to second order, the tensor-to-scalar ratio up to third order, and the spectral indices and running converge to all orders. This indicates that the expansion scheme is incompatible with the working equations derived from UA for the power spectra but compatible with that of the spectral indices. For those quantities that involve logarithmically divergent terms in the higher-order parts, existing results in the literature for the convergent lower-order parts calculated in the equivalent fashion should be viewed with some caution; they do not rest on solid mathematical ground.}


\begin{document}
\maketitle

\bigskip
\bigskip
\section{Introduction}
\label{intro}
Cosmological inflation \cite{Starobinsky:1979ty, Guth:1980zm, Linde:1981mu, Albrecht:1982wi, LiddlenLyth, Mukhanov:2005sc, Martin:2014, Albrecht:2002uz} is a rapid exponential expansion of spacetime. It stretched the quantum fluctuations which eventually gave birth to large-scale structures that we nowadays observe as galaxies and clusters of galaxies. Being able to provide an elegant solution to the homogeneity and flatness problems associated with fine-tuning the appropriate initial conditions for the Universe and explain the emergence of the seeds of structure formation, it serves as a good complementary patch for the standard Big Bang Cosmology. What we have today is a paradigm with great explanatory power involving a hybrid theory composed of inflation and Big Bang theory.

Owing to its incomplete nature however, inflationary cosmology is not without hurdles. When making predictions or deriving expressions for physical quantities that we want to compare with experiments, one is faced with at least two possible problems. The first one deals with the possible disagreement of the prediction with experiment. The second one involves the self-consistency of the prediction. Focusing on the latter, one may consider for instance, the possible emergence of divergent quantities that need to be regularised or renormalised (see for instance, Refs. \cite{Parker:2007,Bastero-Gil:2013,Alinea:2015pza}), contradictory ends of calculations of non-Gaussianity (see Refs. \cite{Bartolo:2004if} for a review of non-Gaussianity), or some mathematical artifacts that when scrutinized may lead to misleading results.

We take the last point above seriously in examining the calculation of the power spectrum among others, carried out in Ref. \cite{Martin:2013} (see also Refs. \cite{Lorenz:2008et,Zhu:2014wfa,Zhu:2014aea}), using the method of the uniform approximation (UA) \cite{Olver,Miller:2006}. Our inquiry centres around the possible emergence of logarithmic divergences as mathematical artifacts of one calculation scheme for relevant physical quantities in inflationary cosmology. To elaborate, we start with the Mukhanov-Sasaki equation for scalar perturbations in the inflation model we are interested in this work namely, slow-roll $ k $-inflation \cite{ArmendarizPicon:1999rj,Garriga:1999vw}. It states
\begin{align}
    v''_k 
    +
    \left(k^2 c_s^2 - U_s\right)v_k
    =
    0,
    \label{scalarMuk}
\end{align}
where $ v_k $ is related to the gauge-invariant cosmological perturbation $ \mathcal R_k $ \cite{Bardeen:1983} as $ v_k \equiv z\mathcal R_k $ with $ z^2 \equiv 2a^2\epsilon /c_s^2 $ and $a$ is the scale factor, $ \epsilon \equiv -\dot H/H^2 $ is the first slow-roll parameter defined here in terms of the Hubble parameter  $ H $, $ k $ is the wavenumber, $ c_s $ is the speed of sound, $ U_s $ is the effective potential, and the symbols dot and prime denote differentiation with respect to the coordinate time $t$ and conformal time $ \eta  $ respectively. The effective potential takes the form $ U_s = (1/\eta ^2) (2 + \cdots) $, where the dots from the point of view of slow-roll $ k $-inflation, represent quantities that are small in comparison to the addend two. This form of the potential is suggestive of a perturbative solution. Furthermore, owing to the structure of the Mukhanov-Sasaki equation pertaining to its behaviour for small- and large-wavelength modes, it is subject to semi-analytical approaches to solving $ v_k $ and hence, calculation of relevant quantities like the power spectrum.

The UA is one of these semi-analytical approaches, that involves using a \textit{single} approximating function to find the solution of a given differential equation. Although this method is quite well known in mathematics as part of the broad area called \textit{asymptotic analysis} \cite{Olver,Miller:2006}, it appears to have only found its way into solving the Mukhanov-Sasaki equation in 2002 through the work in Ref. \cite{Habib:2002yi}. The authors mainly dealt with demonstrating the method in calculating the power spectra and spectral indices for the slow-roll single-field inflation models and its variants. Specifically, they laid down the working equations for these quantities to leading order in the UA and highlighted the advantages of the method namely: (a) systematically improvable, (b) has definite error bounds, and (c) does not rely on slow-roll parameters being constant. It was later extended and given a more detailed account in the follow-up paper where the working equations for the power spectra and spectral indices were derived up to next-to-leading order in the UA \cite{Habib:2004kc}. The natural progression\footnote{Along the way, other areas of investigation were pursued by several authors. These include studies involving multiple turning points \cite{Zhu:2013upa}, gravitational quantum effects modifying the dispersion relation corresponding to the Mukhanov-Sasaki equation \cite{Zhu:2014aea}, chaotic $\lambda\phi^4$ inflation \cite{Rojas:2011rg}, etc.} after this was the calculation of the power spectra in the more general $ k $-inflation where unlike that of the canonical single-field inflation, the speed of sound is not constant. In Ref. \cite{Lorenz:2008et} the power spectra were calculated up to first order in terms of the so-called \textit{Hubble} and \textit{sound flow parameters} (see Subsec. \ref{hubbleAndSound} and \ref{seriesExpandOther}). In a follow-up paper, the power spectra were computed up to second order in terms of the same parameters \cite{Martin:2013}. Note that these last two calculations were done to the lowest order of precision in the UA, that is, up to leading order, but up to second order in the Hubble and sound flow parameters. Up to second order in the Hubble and sound flow parameters, the computations for the higher-order corrections in the UA for the power spectra were carried out in Ref. \cite{Zhu:2014wfa}, superseding the results in Ref. \cite{Martin:2013}.

The problem that we consider in this work does not involve higher-order corrections in the UA that were considered in Ref. \cite{Zhu:2014wfa}. Restricting to leading order in the UA, our main focus is on the issue of calculating physical quantities to any order in terms of the Hubble and sound flow parameters. Consider for instance, the power spectrum for scalar perturbations. One has $P^{(e)}_s = P_s(1 + \cdots)$, where $P^{(e)}_s$ is the exact power spectrum and the dots represent higher-order corrections in the UA. The term $P_s$ is the base quantity we are interested in, that can be computed in terms of the Hubble and sound flow parameters. Since there is no practical limitation as to how far the ``dots'' can be computed, this base quantity $P_s$ limits up to what order in terms of the Hubble and sound flow parameters $P^{(e)}_s$ can be calculated. Furthermore, any defect in the value of $P_s$ is correspondingly reflected in the approximation of the exact power spectrum. Now, based on Ref. \cite{Martin:2013}, one has
\begin{align}
	P_s 
	&=
	-\lim_{\eta \rightarrow 0^-}
	\frac{k^3}{4\pi ^2}
	\frac{\eta c_s^2}{2a^2\epsilon _1 \nu _s}
	e^{2\Psi _s}
	\label{scalarP}
\end{align}
where 
\begin{align}
	\Psi _s
	&=
	\int_{\eta _*}^{\eta }
	\dd\tau \,
	\sqrt{\frac{\nu^2 _s}{\tau^2} - k^2c_s^2},
	\quad
	\nu ^2_s
	\equiv
	\frac{1}{4} + \eta ^2 U_s,
	\label{psiS}
\end{align}
with $ \epsilon _1 = \epsilon$, so-renamed to fit the definitions of Hubble and sound flow functions (see Subsec. \ref{hubbleAndSound}) and $ \eta _* $ is the value of the conformal time at the so-called \textit{turning point} (see Subsec. \ref{uniformApprox}). The natural way to compute $ P_s $ as was done in Refs. \cite{Lorenz:2008et,Martin:2013}, is to expand the set of quantities $c_s^2$, $a^2$, $\epsilon_1$, and $\nu_s$ in the coefficient of the exponential function in (\ref{scalarP}) and the integrand in (\ref{psiS}) about the turning point. Consider for instance, $ \epsilon _1 $: 
\begin{align}
	\epsilon _1
	&=
	\epsilon _{1*} - \epsilon _{1*}\epsilon _{2*}\ln \frac{\eta }{\eta _*}
	-
	\epsilon _{1*}^2\epsilon _{2*} \ln \frac{\eta }{\eta _*}
	-
	\frac{1}{2}\epsilon _{1*}\epsilon _{2*}^2 \ln^2 \frac{\eta }{\eta _*}
	+
	\cdots.
\end{align}
Here, $\epsilon_{n*}$ denotes the Hubble flow function ($\epsilon_n$) evaluated at the turning point. Similar expressions can be derived for other quantities involved in the expression for $ P_s $ above (see Sec. \ref{seriesExpand}). The characteristic form is a chain of Hubble ($\epsilon_n$) and sound ($\delta_n$) flow \textit{functions}  evaluated at the turning point---the Hubble and sound flow \textit{parameters} ($ \epsilon _{n*}, \delta _{n*} $)---and an increasing power of the logarithm of the conformal time. Note that the expansion above for $ \epsilon _1 $ and other quantities such as $ a $, $ \nu _s $, etc, can be carried out indefinitely with respect to ($ \epsilon _{n*}, \delta _{n*} $) (see Subsec. \ref{seriesExpandOther}). One then performs the integration in (\ref{psiS}) and long algebraic manipulations follow in an attempt to find $ P_s $ to a desired order with respect to ($ \epsilon _{n*}, \delta _{n*} $).

As already mentioned, in Ref. \cite{Lorenz:2008et} $ P_s $ was calculated up to first order while in Ref. \cite{Martin:2013}, it was calculated up to second order with respect to ($ \epsilon _{n*}, \delta _{n*} $).\footnote{In Ref. \cite{Zhu:2014wfa}, the power spectra were also calculated up to second order with respect to ($ \epsilon _{n*}, \delta _{n*} $) but all other quantities including spectral indices, running, and tensor-to-scalar ratio were computed at least to third order in ($ \epsilon _{n*}, \delta _{n*} $).}  We point out that these calculations rely heavily on the striking balance of the logarithmic terms, that is, those involving $\ln(\eta/\eta_*)$, in the exponential term and in its literal coefficients in the expression for $ P_s $ in (\ref{scalarP}). Indeed, the fact that these logarithmic terms completely cancel out for the first- and second-order calculations allows one to take the limit as $ \eta \rightarrow 0^- $ in (\ref{scalarP}) and finally come up with a finite result for $ P_s $. In Ref. \cite{Martin:2013}, such an exact cancellation is considered as a ``consistency check'' of the method that they used because the power spectra involve constant modes. This seems to give us no reason to suspect that the calculation of the power spectra and other physical quantities such as spectral index and tensor-to-scalar ratio can be carried out indefinitely to any order with respect to $ (\epsilon _{n*}, \delta _{n*}) $.

Our aim in this paper is to investigate how far one can go with such a calculation laid down in Ref. \cite{Martin:2013} before possibly hitting a divergence, thereby, check its consistency when applied to the calculation of the higher order parts of several physical quantities with respect to $ (\epsilon _{n*}, \delta _{n*}) $, and characterise the nature of the (possible) divergence(s). In particular, we consider the calculation of the power spectra, spectral indices, running of the spectral index for scalar and tensor perturbations, and tensor-to-scalar ratio, using the UA and the expansion scheme employed in Ref. \cite{Martin:2013}\footnote{An equivalent expansion scheme was also used in the calculations in Ref. \cite{Zhu:2014wfa}.}. We wish to find out up to what order in terms of ($ \epsilon _{n*}, \delta _{n*} $) one can calculate these quantities without encountering a logarithmic divergence in the end result. We note at this point that although current experiments may not call for predictions beyond the second order with respect to ($ \epsilon _{n*}, \delta _{n*} $), our aim here is more geared towards mathematical consistency. The hope is to point out limitations in the current computational scheme to give way for possible corrections and self-consistent predictions. 

This paper is organised as follows. In the next section, we discuss the set up of this work. Specifically, we cover the type of inflation we are interested in namely, $ k $-inflation, and then discuss the Hubble and sound flow functions and the UA. At the end of this section, we state the expression for the power spectra for the scalar and tensor perturbations to lowest order of precision in the UA. In Sec. \ref{seriesExpand}, we derive a series expansion of the conformal time in terms of the Hubble and sound flow functions. The existence of such a series as we shall see, plays a significant role in calculating the series expansion of the Hubble and sound flow functions with respect to the conformal time and their values at the turning point. With the working equations for the power spectra and the expansion for the Hubble and sound flow functions in hand, in Sec. \ref{logDiv} we push the limit of the currently available results for the power spectra and identify at what order the logarithmic divergences start to appear. Then in Sec. \ref{specRunTensor} we consider other physical quantities such as the spectral index, running, and the tensor-to-scalar ratio. Having laid down all the necessary expressions for physical quantities such as the power spectra, tensor-to-scalar ratio, etc., we make detailed remarks in Sec. \ref{logDivRamification} on the origin of logarithmic divergences and their ramifications in light of existing calculations in the literature. In the last section, we state our conclusion and future prospects in relation to the possibility of curing the logarithmic divergences that we encounter in this work.

\bigskip
\section{Set up}
\label{model}

\bigskip
\subsection{$ k $-inflation}
\label{kInflation}
We consider a scalar field $ \phi $ called the \textit{inflaton}, that is minimally coupled to the background geometry described by Einstein gravity. The most general local action involving an arbitrary function $ P(X, \phi ) $ where $ X \equiv -\frac{1}{2} g^{\mu\nu}\nabla_\mu\phi \nabla_\nu \phi $ can be written as
\begin{align}
    S
    &=
    \frac{1}{2}
    \int \sqrt{-g}\, \dd^4 x
    \big[
        M_\text{Pl}^2\, R
        +
        2P(\phi,\, X)
    \big].
    \label{actionS}
\end{align}
Here, $ R $ is the Ricci scalar, $ g $ is the determinant of the metric $g_{\mu\nu}$, and $ M_\text{Pl} $ is the reduced Planck mass that we hereafter, take (the square) as unity for brevity. Such an action corresponds to the so-called \textit{k-inflation}. \cite{ArmendarizPicon:1999rj,Garriga:1999vw} The scope of this work involves slow-roll $ k $-inflation. When $ P = X - V(\phi) $, where $ V(\phi) $ is the inflaton potential, the $ k $-inflationary action above reduces to that of the canonical single field inflation.

The background geometry corresponding to the action above is described by the conformally flat Friedmann-Lema\^itre-Robertson-Walker (FLRW) metric given by
\begin{align}
	\dd s^2
	&=
	\dd t^2
	-
	a^2(t)\delta _{ij}\dd x^i \dd x^j.
\end{align}
With this metric and the equation for the action given above, the Friedmann and continuity equations take the form
\begin{align}
    H^2
    &=
    \frac{8\pi G}{3}E, 
    \nonumber
    \\[0.5em]
    \dot E
    &=
    -3{H}\left(E+P\right), 
    \label{friedmanContinuity}
\end{align}
respectively, where $ E $ is the energy density, $ H \equiv \dot a/a$ is the Hubble constant, and an overdot denotes differentiation with respect to the coordinate time $ t $.

To study the perturbations about our otherwise, conformally flat background, we must choose appropriate slicing and threading of spacetime. In connection to this, we find it convenient to use the Arnowitt-Deser-Misner (ADM) formalism \cite{Arnowitt:1959ah} and decompose the (perturbed) metric as
\begin{align}
	\dd s^2	
	&=
	N^2 \dd t^2
	-
	h_{ij}(\dd x^i + N^i\dd t)(\dd x^j +  N^j\dd t),
\end{align}
where $ N $ and $ N^i $ are the lapse and shift functions respectively. Furthermore, we choose the \textit{uniform inflaton gauge} (also called the \textit{comoving gauge}) where the inflaton fluctuation vanishes, \textit{i.e.}, $ \delta \phi = 0 $, and the spatial components of the metric take the form 
\begin{align}
    g_{ij}
    &=
    a^2(\eta)e^{2\mathcal R}\big(e^\gamma\big)_{ij},
    \qquad
    (\partial^i\gamma_{ij} = \gamma^{i}{}_i = 0).
\end{align}
Here, $ \mathcal R $ and $ \gamma_{ij} $ are the gauge-invariant scalar and tensor perturbations respectively. The tensor perturbations satisfy the auxiliary conditions that they be both transverse and traceless.

It is now possible to write down the equations of motion for the perturbations $ \mathcal R $ and $ \gamma _{ij} $. For the scalar perturbation, one decomposes the action as $ S = S^{(0)} + S^{(2)} + S^{(3)} + \cdots$, where $ S^{(n)} $ means the $ n $th-order action with respect to $ \mathcal R $ (see e.g., Ref. \cite{Chen:2010xka}). The background corresponds to the zeroth-order part while the interaction starts at $ n = 3 $. It is from the free part $ S^{(2)} $ that we derive our equations of motion. In Fourier space, we find the so-called \textit{Mukhanov-Sasaki equation} given as
\begin{align}
    v''_k 
    +
    \left(k^2 c_s^2 - \frac{z''}{z}\right)v_k
    =
    0,
    \label{scalarMukii}
\end{align}
where $ v_k $ is related to $ \mathcal R_k $ by $ v_k = z\mathcal R_k $, the quantity $ k $ is the wavenumber, $ z^2 \equiv 2a^2\epsilon/c_s^2 $ with $ \epsilon \equiv -\dot H/H^2$ being the first slow-roll parameter, and the symbol prime indicates differentiation with respect to the conformal time $ \eta $ with $ \dd\eta \equiv \dd t/a $. The speed of sound $ c_s^2 $ is defined as
\begin{align}
    c_s^2
    \equiv
    \frac{P_{,X}}{E_{,X}}
    =
    \frac{P_{,X}}{P_{,X}+2XP_{,XX}},
    \label{soundspeed}
\end{align} 
where the subscript $ X $ denotes differentiation with respect to $ X $. The second equality follows from the relation $ E = 2XP_{,X} - P$ which is due to 00-component of the energy-momentum tensor ($ T^0{}_0 $) being identified as the energy density. For stable solutions wherein $c_s^2 > 0$, corresponding to a bounded total Hamiltonian involving both perturbations and background, $P_{,X} > 0$ and $2X P_{,XX}+P_{,X} > 0$ (see Refs. \cite{ArmendarizPicon:1999rj, Garriga:1999vw, Bruneton:2006gf} for more details). Note in (\ref{soundspeed}) that when $ P = X - V $ we have $ P_{,XX} = 0$ giving $ c_s^2 = 1 $, and we recover the usual Mukhanov-Sasaki equation for the canonical single-field inflation. For the tensor perturbation, following a similar route, we find
\begin{align}
    \mu ''_k 
    +
    \left(k^2 - \frac{a''}{a}\right)\mu_k
    =
    0.
    \label{tensorMuk}
\end{align}
where $ \mu_k/a $ is the amplitude of gravitational waves.

Given the two equations of motion above, the (dimensionless) power spectra for scalar ($ P_s^{(e)} $) and tensor ($ P_t^{(e)} $) perturbations can be written as
\begin{align}
	P_s^{(e)}
	&=
	\frac{k^3}{2\pi ^2}
	\left|
		\frac{v_k}{z}
	\right|^2,
	\qquad
	P_t^{(e)}
	=
	\frac{2k^3}{\pi ^2}
	\left|
		\frac{\mu_k}{a}
	\right|^2,
	\label{powerSpec}
\end{align}
where $(e)$ means exact. It then remains to find the respective solutions (or approximate solutions) of the two differential equations above for $ v_k $ and $ \mu_k $ to determine $ P_s \approx P^{(e)}_s $ and $ P_t \approx P_t^{(e)} $, where $P_s$ and $P_t$ represent approximations based on semi-analytical method such as the UA.

\bigskip
\subsection{Hubble and Sound Flow Functions}
\label{hubbleAndSound}
When $ c_s^2 = 1 $, a closed-form solution exists for the power law inflation \cite{Lucchin:1984yf}. In such a case, $ \epsilon = \text{constant}$ and the equations of motion for scalar and tensor perturbations take exactly the same form. In general, however, the two equations of motion above do not have a closed-form solution.\footnote{A short review of the known analytical solutions can be found in Ref. \cite{Martin:2000ei}. The authors of this work also presented their own additional analytical solutions.} Nevertheless, the structure of the Mukhanov-Sasaki equation makes it amenable to semi-analytical calculation such as the UA---the method we employ in this work (see Subsec. \ref{uniformApprox}).

To proceed with the calculation, one introduces a set of functions or parameters in terms of which, physical quantities like power spectra may be expressed. The most commonly used is the set of slow-roll parameters defined (albeit, slightly different from each other) in Refs. \cite{Stewart:1993bc,Gong:2001he,Lidsey:1995np,Martin:1999wa}. In Ref. \cite{Schwarz:2001vv} a new set of functions called \textit{horizon} (or \textit{Hubble}) \textit{flow functions} was introduced. As the authors pointed out, it offers the advantages of being (a) model-independent, (b) concise, and (c) easily memorised. All the slow-roll parameters or functions defined in Refs. \cite{Stewart:1993bc,Gong:2001he,Lidsey:1995np,Martin:1999wa} can be written in terms of these new functions. Additional flow functions called the \textit{sound flow functions} were introduced in Ref. \cite{Lorenz:2008et} to include the speed of sound in the calculation of the power spectra for $ k $-inflation. In this work, owing to the advantages mentioned above, we make use of these two sets of functions.

The dimensionless Hubble flow functions ($ \epsilon _n $) and sound flow functions ($ \delta _n $) are given by the recursive definitions \cite{Schwarz:2001vv,Martin:2013}
\begin{align}
    \epsilon_{n+1}
    &\equiv
    \frac{\dd\ln{\epsilon_n}}{\dd N},
    \qquad
    \epsilon_0 
    \equiv 
    \frac{H_{i}}{H},
    \label{hubbleFlow}
    \\[0.5em]
    \delta_{n+1}
    &\equiv
    \frac{\dd\ln{\delta_n}}{\dd N},
    \qquad
    \delta_0 
    \equiv
    \frac{c_{s,i}}{c_s},
    \label{soundFlow}
\end{align}
where $N \equiv \int \dd a/a $ is the number of $ e $-folds, the number $ n $ is a non-negative integer, and the subscript $ i $ indicates some initial value. Note that with the above definitions, $ \epsilon _1 $ is simply the usual slow-roll parameter $ \epsilon \equiv -\dot H/H^2 $. Furthermore, $ \delta _1 $ serves as a measure of the relative rate of change of the speed of sound; \textit{i.e.}, $ \delta _1 = -(1/H)(\dot c_s/c_s) $.

With the Hubble and sound flow functions in hand, it is now straightforward to rewrite the Mukhanov-Sasaki equations for scalar and tensor perturbations. The only relevant parts that need modification are the terms $ z''/z $ and $ a''/a $. We identify these as the effective potentials $ U_s $ and $ U_t $,
\begin{align}
	U_s
	=
	\frac{z''}{z},
	\qquad
	U_t
	=
	\frac{a''}{a},
	\label{effPot}
\end{align}
in analogy with the equation for a parametric oscillator. With these identifications and the recursive definitions above for $ \epsilon _n $ and $ \delta _n $ we find 
\begin{align}
    U_s (\eta)
    &=
    (aH)^2 \Big[2-\epsilon_1 
    +
    \tfrac{3}{2}\,\epsilon_2
    +
    \tfrac{1}{4}\,\epsilon_2^2 
    -
    \tfrac{1}{2}\epsilon_1\epsilon_2
    +
    \tfrac{1}{2}\,\epsilon_2 \epsilon_3 
    +
    \big(3-\epsilon_1 +\epsilon_2\big)\delta_1
    +
    \delta^2_1 +\delta_1 \delta_2 \Big],
    \nonumber
    \\[0.5em]
	U_t (\eta)
    &=
    (aH)^2(2-\epsilon_1)
    \label{effU}
\end{align}
These two equations for the effective potentials are \textit{exact}. The structure of the expressions for the potentials tells us that for the usual slow-roll inflation where $ c_s^2 = 1 $, owing to the fact that $ \eta = -1/(aH)(1 + \epsilon _1 + \cdots) $, the solutions of the Mukhanov-Sasaki equations can be well approximated by Hankel functions \cite{Stewart:1993bc} when the slow-roll parameters are nearly constant. Beyond this is a generalization to the case where the slow-roll parameters are not constant, inclusion of higher-order corrections in terms of these parameters, and a consideration of non-constant speed of sound.  The important corrections may be captured by semi-analytical approaches utilizing Green's function \cite{Gong:2001he,Gong:2004kd}, WKB approximation \cite{Martin:2002vn,Casadio:2005xv,Casadio:2005em}, the UA \cite{Habib:2004kc,Martin:2013,Zhu:2014wfa}, etc.

\bigskip
\subsection{Uniform Approximation and Power Spectra}
\label{uniformApprox}
From (\ref{scalarMuk}) and the first of (\ref{effPot}), the Mukhanov-Sasaki equation for scalar perturbations in the canonical single-field inflation takes the form 
\begin{align}
	v_k''
	+
	\left(
		k^2 - U_s
	\right)v_k
	&=
	0;
	\qquad
	U_s \equiv z''/z
\end{align}
When the potential term is negligible in comparison to the wavenumber, the solution of this differential equation is simply a sum of two sinusoids. On the other hand, for long-wavelength modes, the differential equation approaches a form given by $ v_k'' - (z''/z) v_k = 0 $. It has the obvious solution $ v_k \sim z $ which corresponds to the frozen modes $ \mathcal R_k \sim \text{constant} $ involved in the power spectrum. More formally, for subhorizon modes wherein $ |k/(aH)| \sim |k\eta| \gg 1  $, we have a sinusoidal solution and for superhorizon modes wherein $ |k/(aH)| \sim |k\eta| \ll 1 $, we have $ v_k\sim z $ corresponding to $ \mathcal R_k \sim \text{constant}$.

In the case of the more general $ k $-inflation---the type of inflation we are interested here---the effective potential depends on a chain of Hubble and sound flow functions (cf. (\ref{effU})) and as pointed out earlier, the resulting differential equation may not have a simple closed-form solution. However, the behaviour of the solution at extremal ends of the spectrum of $ k\eta \sim k/(aH) $ remains the same; that is, we have two solutions corresponding to subhorizon and superhorizon modes and these two are connected somewhere between the two extreme values of $ k\eta $. In principle, one can do a sort of matching typical of a square-well potential problems in one-dimensional quantum mechanics. In contrast, such a sort of matching different solutions for different regions is not explicitly existent in the UA. Instead, there is a \textit{single} approximating solution that takes on all the burden of adjustment to fit the boundary or initial conditions and the differential equation that it satisfies.

In the case of the Mukhanov-Sasaki equation, the ``differing'' behaviour of the solution at extremal ends in the range of values of $ k\eta $ is taken into account by considering the turning-point. More specifically, one can rewrite the Mukhanov-Sasaki equation for scalar perturbation in the canonical single-field inflation as\footnote{Analogous expressions hold for the tensor perturbations.}
\begin{align}
	v_k''
	+
	\left[
		\frac{1}{4\eta ^2} 
		-
		g_s(\eta )
	\right]
	=
	0,
\end{align}
where
\begin{align}
	g_s(\eta )
	&=
	\frac{\nu_s ^2}{\eta ^2} - k^2;
	\qquad
	\nu_s^2 \equiv \frac{1}{4} + \eta ^2U_s.
\end{align}
The turning point corresponds to the value $ \eta = \eta _*$ at which $ g_s(\eta_*) = 0 $ and the solution is in transition from one sort of behaviour to another; say, from oscillatory to decaying. Fortunately, owing to the range of values of $ \eta $ being non-positive, the Mukhanov-Sasaki equation has only one turning point. With this feature, it turns out that the solution to leading order in the UA can be written in terms of Airy functions. More specifically, one takes
\begin{align}
	v_k
	=
	\lim_{k\eta \rightarrow 0^{-}}
	\bigg[
		c_1\left(\frac{f}{g}\right)^\frac{1}{4}
		\text{Ai($ f $)}
		+
		c_2\left(\frac{f}{g}\right)^\frac{1}{4}
		\text{Bi($ f $)}		
	\bigg],
	\label{vkLowest}
\end{align}
where $ c_1 $ and $ c_2 $ are constants and 
\begin{align}
	f(\eta )
	&=
	\frac{|\eta - \eta _*|}{\eta - \eta _*}\bigg|
		\frac{3}{2}\int _{\eta _*}^\eta \dd\tau \sqrt{g(\tau )}
	\bigg|^\frac{2}{3}.
	\label{eff}
\end{align}
As one can see, in the UA, the problem of solving the Mukhanov-Sasaki equation is converted to solving integrals that serve as arguments of the Airy functions in (\ref{vkLowest}). The ability to find $ v_k $ in closed form to some orders with respect to $ (\epsilon _{n*}, \delta _{n*}) $ then relies on our being able to calculate those integrals in closed form.\footnote{The error corrections in UA also take an integral form; see for instance, Refs. \cite{Habib:2004kc,Zhu:2014wfa}.}

In the calculation of power spectra and other physical quantities like spectral tilt and tensor-to-scalar ratio, we are only interested in superhorizon modes so we take $ v_k \sim \lim_{k\eta \rightarrow 0^-} (f/g)^\frac{1}{4}\text{Bi}(f) $. For large $ f $ corresponding to $ k\eta \rightarrow 0^- $, we have the asymptotic form $ \text{Bi}(f)\sim e^{\frac{2}{3}f^{\frac{3}{2}}}/(\sqrt{\pi} f^\frac{1}{4})$. It follows that
\begin{align}
	v_k
	&\sim
	\lim_{k\eta \rightarrow 0^-}
	\frac{e^{\frac{2}{3} f^\frac{3}{2}}}{g^{\frac{1}{4}}},
	\label{vkApp}
\end{align}

The generalisation to $ k $-inflation is quite straightforward. Effectively, one simply replaces the wavenumber $ k $ by $ k\,c_s $ so that the function $ g $ is modified as
\begin{align}
	g_s(\eta )
	&=
	\frac{\nu _s^2}{\eta ^2}
	-
	k^2 c_s^2
	\label{gScalar}
\end{align}
The function $ f $ and $ v_k $ given by (\ref{eff}) and (\ref{vkApp}) respectively, take the same form except that they involve the more general expression for $ g_s $ given above. Assuming that the initial state of the perturbations is the Bunch-Davies vacuum as in Ref. \cite{Martin:2013}, the power spectrum given in the first of (\ref{powerSpec}) to leading order in the UA can be rewritten as
\begin{align}
	P_s 
	&=
	-\frac{k^3}{8\pi ^2}
	\frac{\eta c_s^2}{a^2 \epsilon _1 \nu _s}
	e^{2\Psi _s}
	\label{scalarPii}
\end{align}
where 
\begin{align}
	\Psi _s
	&=
	\int_{\eta _*}^{\eta }
	\dd\tau \,
	\sqrt{\frac{\nu^2 _s}{\tau^2} - k^2c_s^2},
	\quad
	\nu ^2_s
	\equiv
	\frac{1}{4} + \eta ^2 U_s,
\end{align}
For the tensor perturbations, the power spectrum given by the second of (\ref{powerSpec}) can be approximated as
\begin{align}
	P_t
	&=
	-\frac{2k^3}{\pi ^2}
	\frac{\eta }{a^2 \nu _t}
	e^{2\Psi _t},
	\label{tensorP}
\end{align}
where
\begin{align}
	\Psi _t
	&=
	\int_{\eta _*}^{\eta }
	\dd\tau \,
	\sqrt{\frac{\nu^2 _t}{\tau^2} - k^2},
	\qquad
	\nu^2 _t
	\equiv
	\frac{1}{4} + \eta ^2 U_t.
\end{align}
Note that the limit $ \eta \rightarrow 0^- $ is implicit in the expressions for $ P_s $ and $ P_t $ above. These two expressions for the power spectra are our working equations to leading order in the UA.

\bigskip
\bigskip
\section{Series Expansion of the Conformal time and the Hubble and Sound Flow Functions}
\label{seriesExpand}

\bigskip
\subsection{Conformal Time}
\label{seriesExpand2}
With the working equation for $ P_s $ and $ P_t $ given in the previous section, the possibility of calculating power spectra to \textit{any order} in the Hubble and sound flow parameters mainly relies on the convenience provided by the recursive definitions of $ \epsilon _n $ and $ \delta _n $ given by (\ref{hubbleFlow}) and (\ref{soundFlow}) and the ability to express  the conformal time in terms of $ \epsilon _n $ up to any order with respect to $ \epsilon _n $. It is often the case that recursive integration is used to expressed $ \eta $ up to second or third order in $ \epsilon _n $ with the possibility of extending it indefinitely. In this subsection, we give a short and simple derivation of a series expansion for the conformal time by indefinitely extending the recursive integration by parts usually used in the literature.

We start with the definition $ \dd\eta \equiv \dd t/a $. With $ \eta$ in the range $ (-\infty, 0] $, this definition implies that
\begin{align}
	\eta 
	&=
	\int \frac{\dd a}{a^2 H}
	=
	-\frac{1}{aH}
	+
	\int \frac{\dd t}{a}\, \frac{\dd}{\dd t}\big(H^{-1}\big).
\end{align}
The quantity $ \dd t/a = \dd a/(a^2H) $ appears again in the integrand in the right hand side so following the first one, we perform another round of integration by parts.
\begin{align}
	\eta 
	&=
	-\frac{1}{aH} 
	-
	\frac{1}{a}H^{-1}
	\frac{\dd}{\dd t}\big(H^{-1}\big)
	+
	\int \frac{\dd t}{a}
	\frac{\dd }{\dd t}\bigg(
		H^{-1}\frac{\dd}{\dd t}\big(H^{-1}\big)
	\bigg)
\end{align}
Continuing, we find
\begin{align}
	\eta 
	&=
	\sum_{n = 0}^{\infty}\tau _n
	=
	-\frac{1}{a}
	\sum_{n = 0}^{\infty}\bigg(
		H^{-1}
		\frac{\dd}{\dd t}
	\bigg)^n H^{-1},
	\label{seriesEta}
\end{align}
where
\begin{align}
	\tau _n
	&=
	-\frac{1}{a}
	\bigg(
		H^{-1}
		\frac{\dd}{\dd t}
	\bigg)^n H^{-1}.
	\label{tauN}
\end{align}
Equation (\ref{seriesEta}) is our sought-for series expansion of the conformal time and our main result in this subsection. It is essentially a translation from the relatively more incommodious language of integrals to a convenient language of differentiation. To verify its validity, one can simply differentiate $\eta$ with respect to $t$ and show that $ \dd \eta/\dd t = 1/a $ in accord with the definition $ \dd\eta \equiv \dd t/a $.

The infinite sum above for the conformal time is a Neumann series \cite{Lebedev:1996} of the operator $ H^{-1} \dd/\dd t$ acting on $ H^{-1} $. 
\begin{align}
	\eta 
	&=
	-\frac{1}{a}\bigg[
		1 + \bigg(H^{-1}\frac{\dd}{\dd t}\bigg)
		+
		\bigg(H^{-1}\frac{\dd}{\dd t}\bigg)^2
		+
		\cdots
	\bigg]
	H^{-1}
	\nonumber
	\\[0.5em]
	\eta 
	&=
	-\frac{1}{a}\bigg(
		1 - H^{-1}\frac{\dd}{\dd t}
	\bigg)^{-1}H^{-1}.
\end{align}
As such, had we started with 
\begin{align}
	-\bigg(
		1 - H^{-1}\frac{\dd}{\dd t}
	\bigg)a\eta 
	=
	H^{-1}
	\quad
	\Leftrightarrow
	\quad
	\dd\eta 
	=
	\frac{\dd t}{a},
	\qquad
	(a \ne 0,\, H^{-1} \ne 0)
\end{align}
we could have ended with the same expansion for $ \eta  $  given by (\ref{seriesEta}) without going through recursive integration.

Using the series expansion for $ \eta $ given above and the recursive definitions for the Hubble flow functions we find
\begin{align}
	\bar \tau _0
	&=
	1
	\nonumber
	\\[0.5em]
	\bar \tau _1
	&=
	\epsilon _1
	\nonumber
	\\[0.5em]
	\bar \tau _2
	&=
	\epsilon _1^2 + \epsilon _1\epsilon _2
	\nonumber
	\\[0.5em]
	\bar \tau _3
	&=
	\epsilon _1^3 + 3\epsilon _1^2\epsilon _2
		+ 
		\epsilon _1\epsilon _2^2
		+
		\epsilon _1\epsilon _2\epsilon _3
	\nonumber
	\\[0.5em]
	\bar \tau _4
	&=
	\epsilon _1^4
	+
	6\epsilon _1^3\epsilon _2
	+
	7\epsilon _1^2\epsilon _2^2
	+
	\epsilon _1\epsilon _2^3
	+
	4\epsilon _1^2\epsilon _2\epsilon _3
	+
	3\epsilon _1\epsilon _2^2\epsilon _3
	+
	\epsilon _1\epsilon _2\epsilon _3^2
	+
	\epsilon _1\epsilon _2\epsilon _3\epsilon _4
	\nonumber
	\\[0.5em]
	\vdots
\end{align}
where $ \bar \tau _n \equiv -\tau _n(aH) $ is $ \mathcal O(\epsilon ^n) $. Because of the structure of the recursive definition of $ \epsilon _n $, the number of terms involved in $ \bar \tau _n $ grows as $ 2^{n-1} $. With the above expressions for $ \bar \tau _n $, up to third order, the conformal time can be written as
\begin{align}
	\eta 
	&=
	-\frac{1}{aH}\big[
		1 + \epsilon _1\big(
			1 + \epsilon _1 + \epsilon _2
			+
			\epsilon _1^2 + 3\epsilon _1\epsilon _2
			+
			\epsilon _2^2
			+
			\epsilon _2\epsilon _3
		\big)
		+
		\mathcal O(\epsilon^4)
	\big].
	\label{etaThird}
\end{align}
We will use this expression later in an attempt to calculate the power spectrum beyond the second order in $ (\epsilon_{n*}, \delta _{n*})$.

\bigskip
\subsection{Expansion of the Hubble and Sound Flow Functions}
\label{seriesExpandOther}

Reference \cite{Martin:2013} laid down an ingenious expansion scheme to express the quantities involved in the power spectra given by (\ref{scalarPii}) and (\ref{tensorP}) namely, $ a $, $ \nu $, $ c_s^2 $, $ \epsilon _n $, and $ \delta _n $, in terms of $ \eta  $ and $ (\epsilon _{n*},\, \delta _{n*}) $. Since this expansion scheme plays a significant role in the calculation of the power spectrum among others, we present here a thorough review based on the discussion in the mentioned reference. Along the way, we also demonstrate how such a scheme can be used to expressed all the mentioned five quantities above to \textit{any} order\footnote{From hereon, the word \textit{order} pertains to order with respect to the Hubble and sound flow \textit{functions} $ (\epsilon _n, \delta _n) $ or \textit{parameters} $ (\epsilon _{n*}, \delta _{n*}) $ unless otherwise specified, and \textit{not} with respect to the order of precision of the UA. As already mentioned, for the latter, we work only to leading order of precision.}.

The two basic ingredients that we need are (a) the series expansion of the conformal time derived in the previous subsection and (b) the recursive definition of the Hubble and sound flow functions. Proceeding with $ \epsilon _1 $, we start with the relation $ \dot \epsilon _1 = H\epsilon _1 \epsilon _2 $ taken from the definition (\ref{hubbleFlow}). Because $ \dd\eta \equiv \dd t/a $ and $ \eta  $ is as given by expansion (\ref{etaThird}) we find
\begin{align}
	\frac{\dd \epsilon _1}{\dd \eta }
	&=
	-\frac{1}{\eta }\epsilon _1\epsilon _2
	+
	\mathcal O(\epsilon ^3)
\end{align}
Integrating from the turning point at $ \eta = \eta_*$ to some arbitrary value of $ \eta  $ gives
\begin{align}
	\epsilon _1
	&=
	\epsilon _{1*} - \epsilon _{1*}\epsilon _{2*}L
	+
	\mathcal O(\epsilon _*^3),
	\label{prelimEpsilon}
\end{align}
where $ L \equiv \ln (\eta /\eta _*) $.

To go beyond that given by the last equation above for $ \epsilon _1 $, we perform an expansion about the number of $ e $-folds evaluated at the turning point. More specifically, we have
\begin{align}
	\epsilon _1
	&=
	\epsilon _{1*}
	+
	\frac{\dd \epsilon _1}{\dd N}\bigg|_* (\Delta N)
	+
	\frac{1}{2!}
	\frac{\dd^2 \epsilon _1}{\dd N^2}\bigg|_* (\Delta N)^2
	+
	\cdots,
	\label{epsilonExpand}
\end{align}
where $ \Delta N = N - N_* $. Because $ \dd^n \epsilon _1/\dd N^n $ is $ (n + 1) $\textit{th} order, the expansion above for $ \epsilon _1 $ tells us that we only need an expression for $ \Delta N $ up to $ (n-2) $\textit{th} order in order to find an expression for $ \epsilon _1 $ to $ n $\textit{th} order. Needless to say, we only need to find $ \Delta N $ to first order to derive an expansion for $ \epsilon _1 $ one order higher than that of (\ref{prelimEpsilon}).

The preliminary expansion for $ \epsilon _1 $ given by (\ref{prelimEpsilon}) together with the equation for $ \eta  $ given by (\ref{etaThird}) is more than enough to express $ \Delta N $ to first order. In fact, we can go up to second order. To see this, we note from (\ref{etaThird}) that
\begin{align}
	\frac{\dd a}{a}
	&=
	-\frac{\dd\eta }{\eta }\big[
		1 + \epsilon _1 + \epsilon _1^2
		+
		\epsilon _1\epsilon _2
		+
		\mathcal O(\epsilon ^3)
	\big].
\end{align}
Because $ \dd N \equiv \dd a/a $, integration yields
\begin{align}
	\Delta N
	&=
	-\big(
		1 + \epsilon _{1*} 
		+
		\epsilon _{1*}^2
		+
		\epsilon _{1*}\epsilon _{2*}
	\big)L
	+
	\tfrac{1}{2}\,\epsilon _{1*}\epsilon _{2*}L^2
	+
	\mathcal O(\epsilon _*^3),
\end{align}
which is good up to second order as we have just mentioned. When we substitute this in (\ref{epsilonExpand}) we find $ \epsilon _1 $ up to fourth order.
\begin{align}
	\epsilon _1
	&=
	\epsilon _{1*}\big\{
		1 - \epsilon _{2*}L
		-
		\big[
			\epsilon _{1*}\epsilon _{2*}L
			-
			\tfrac{1}{2}\big(\epsilon _{2*}^2
				+
				\epsilon _{2*}\epsilon _{3*}
			\big)L^2
		\big]
		\nonumber
		\\[0.5em]
		&\qquad\quad
		-\,
		\big[
			\big(
				\epsilon _{1*}^2\epsilon _{2*}
				+
				\epsilon _{1*}\epsilon _{2*}^2
			\big)L					
			-
			\big(
				\epsilon _{1*}\epsilon _{2*}\epsilon _{3*}		
				+
				\tfrac{3}{2}\,\epsilon _{1*}\epsilon _{2*}^2
			\big)L^2
			\nonumber
			\\[0.5em]
			&\qquad\qquad\quad		
			+\,
			\tfrac{1}{6}\big(
				\epsilon _{2*}^3
				+
				3\epsilon _{2*}^2\epsilon _{3*}
				+
				\epsilon _{2*}\epsilon _{3*}^2
				+
				\epsilon _{2*}\epsilon _{3*}\epsilon _{4*}
			\big)L^3
		\big]
		+
		\mathcal O(\epsilon_*^4)	
	\big\}
	\label{eps1Expand}
\end{align}
Owing to $ \epsilon _1 $ being in the denominator of the right hand side of (\ref{scalarPii}), this fourth-order expression is just sufficient to calculate the power spectrum $ P_s $ up to third order. 

Following a similar approach, we can also find $ \epsilon _2 $ up to fourth order, 
\begin{align}
	\epsilon _2
	&=
	\epsilon _{2*}\big\{
		1 - \epsilon _{3*}L
		-
		\big[
			\epsilon _{1*}\epsilon _{3*}L
			-
			\tfrac{1}{2}\big(\epsilon _{3*}^2
				+
				\epsilon _{3*}\epsilon _{4*}
			\big)L^2
		\big]
		\nonumber
		\\[0.5em]
		&\qquad\quad
		-\,
		\big[
			\big(
				\epsilon _{1*}^2\epsilon _{3*}
				+
				\epsilon _{1*}\epsilon _{2*}\epsilon _{3*}
			\big)L					
			-
			\big(
				\tfrac{1}{2}\,
				\epsilon _{1*}\epsilon _{2*}\epsilon _{3*}		
				+
				\epsilon _{1*}\epsilon _{3*}^2
				+
				\epsilon _{1*}\epsilon _{3*}\epsilon _{4*}
			\big)L^2
			\nonumber
			\\[0.5em]
			&\qquad\qquad\quad		
			+\,
			\tfrac{1}{6}\big(
				\epsilon _{3*}^3
				+
				3\epsilon _{3*}^2\epsilon _{4*}
				+
				\epsilon _{3*}\epsilon _{4*}^2
				+
				\epsilon _{3*}\epsilon _{4*}\epsilon _{5*}
			\big)L^3
		\big]
		+
		\mathcal O(\epsilon_*^4)	
	\big\},
\end{align}
then $ \epsilon _3 $ up to fourth order, and so on. Once we have these, we can update our expression for $ \Delta N $ up to fourth order, thanks to the infinite series for $ \eta $. Then our expressions for $ \epsilon _n $'s can be updated to higher order. The iterative approach continues and we can express all the Hubble flow \textit{functions} to any desired order in terms of the Hubble flow \textit{parameters} and the conformal time.

We employ the same method for the expansion of the sound flow \textit{functions} with respect to the Hubble and sound flow \textit{parameters} and the conformal time. One starts with the recursive definition given by (\ref{soundFlow}). For $ \delta _1 $ this leads to
\begin{align}
	\delta _1
	&=
	\delta _{1*} - \delta _{1*}\delta _{2*}L
	+
	\mathcal O(\delta _*^3)
\end{align}
Then we use the series
\begin{align}
	\delta _1
	&=
	\delta _{1*}
	+
	\frac{\dd \delta _1}{\dd N}\bigg|_*(\Delta N)
	+
	\frac{1}{2!}\frac{\dd^2\delta _1}{\dd N^2}\bigg|_*(\Delta N)^2
	+
	\cdots.
\end{align}
The same goes for the calculation of other sound flow functions. There is no limit as to the order by which we can express $ \delta _n $ in terms of $ \delta _{n*} $ and $ \epsilon _{n*} $ since $ \Delta N $ can be determined to any desired order. For $ \delta _1 $ and $ \delta _2 $ involved in the effective potential given by the first of (\ref{effU}), we have up to third order, 
\begin{align}
	\delta _1
	&=
	\delta _{1*}\big\{
		1 - \delta _{2*}L
		-
		\tfrac{1}{2}\,\big[
			2\epsilon _{1*}\delta _{2*}L
			-
			\big(
				\delta _{2*}^2
				+
				\delta _{2*}\delta _{3*}
			\big)L^2
		\big]
		+
		\mathcal O\big((\epsilon _*,\delta _*)^3\big)
	\big\}
	\nonumber
	\\[0.5em]
	\delta _2
	&=
	\delta _{2*}\big\{
		1 - \delta _{3*}L
		-
		\tfrac{1}{2}\,\big[
			2\epsilon _{1*}\delta _{3*}L
			-
			\big(
				\delta _{3*}^2
				+
				\delta _{3*}\delta _{4*}
			\big)L^2
		\big]
		+
		\mathcal O\big((\epsilon _*,\delta _*)^3\big)
	\big\}
\end{align}

Of the five quantities ($ a $, $ \nu $, $ c_s^2 $, $ \epsilon _n $ and $ \delta _n $) we mentioned near the beginning of this subsection, we are done with two namely, $ \epsilon _n $ and $ \delta _n $. For the remaining three, we start with the scale factor. Since $ \Delta N = \ln (a/a_*) $ then $ a = a_*e^{\Delta N} $. Because the lowest order in the equation for $ \Delta N $ is $ -L = -\ln (\eta /\eta _*) $ then 
\begin{align}
	a
	&=
	\frac{a_*\eta _*}{\eta }
	e^{\Delta N + L}
	=
	-\frac{e^{\Delta N + L}}{H_*\eta }\big(1 + \epsilon _{1*} + \cdots \big),
\end{align}
where in the second equality, we have used the equation for the conformal time. Squaring this yields
\begin{align}
	a^2
	&=
	\frac{1}{(H_*\eta )^2}\Big\{
		1
		+
		2\big[
			\epsilon _{1*} - \epsilon _{1*}L
		\big]
		+
		\big[
			3\epsilon _{1*}^2
			+
			2\epsilon _{1*}\epsilon_{2*}
			-
			\big(
				6\epsilon _{1*}^2 + 2\epsilon _{1*}\epsilon _{2*}
			\big)L
			\nonumber
			\\[0.5em]
			&\qquad\qquad\quad
			+\,
			\big(
				2\epsilon _{1*}^2
				+
				\epsilon _{1*}\epsilon _{2*}
			\big)L^2
		\big]
		+
		\big[
			4\epsilon _{1*}^3
			+
			8\epsilon _{1*}^2\epsilon _{2*}
			+
			2\epsilon _{1*}\epsilon _{2*}^2
			+
			2\epsilon _{1*}\epsilon _{2*}\epsilon _{3*}
			\nonumber
			\\[0.5em]
			&\qquad\qquad\quad
			-\,
			\big(
				12\epsilon _{1*}^3
				+
				14\epsilon _{1*}^2\epsilon _{2*}
				+
				2\epsilon _{1*}\epsilon _{2*}^2
				+
				2\epsilon _{1*}\epsilon _{2*}\epsilon _{3*}
			\big)L
			\nonumber
			\\[0.5em]
			&\qquad\qquad\quad
			+\,
			\big(
				8\epsilon _{1*}^3
				+
				9\epsilon _{1*}^2\epsilon _{2*}
				+
				\epsilon _{1*}\epsilon _{2*}^2
				+
				\epsilon _{1*}\epsilon _{2*}\epsilon _{3*}
			\big)L^2
			\nonumber
			\\[0.5em]
			&\qquad\qquad\quad
			-\,
			\tfrac{1}{3}\,\big(
				4\epsilon _{1*}^3
				+
				6\epsilon _{1*}^2\epsilon _{2*}
				+
				\epsilon _{1*}\epsilon _{2*}^2
				+
				\epsilon _{1*}\epsilon _{2*}\epsilon _{3*}
			\big)L^3
		\big]
		+
		\mathcal O\big((\epsilon _*,\delta _*)^4\big)
	\Big\}.
	\label{aSq}
\end{align}
The next is the (square of the) speed of sound, $ c_s^2 $. We have
\begin{align}
	c_s^2
	&=
	c_{s*}^2
	+
	\frac{\dd c_s^2}{\dd N}\bigg|_* (\Delta N)
	+
	\frac{1}{2!}
	\frac{\dd^2 c_s^2}{\dd N^2}\bigg|_* (\Delta N)^2
	+
	\cdots.
	\label{csExpand}
\end{align}
By virtue of the recursive definition for $ \delta _n $ given by (\ref{soundFlow}) we find 
\begin{align}
	\frac{\dd \delta _0}{\dd N}
	&=
	\delta _0 \delta _1
	\quad\text{then}\quad
	\frac{\dd c_s^2}{\dd N}
	=
	-2c_s^2 \delta _1,
	\quad
	\frac{\dd^2 c_s^2}{\dd N^2}
	=
	c_s^2\big(
		4\delta _1^2 + 2\delta _1\delta _2
	\big).
\end{align}
It follows that
\begin{align}
	c_s^2
	&=
	c_{s*}^2\Big\{
		1 + 2\delta _{1*}L
		+
		\big[
			2\epsilon _{1*}\delta _{1*}L
			+
			\big(
				2\delta _{1*}^2
				- \delta _{1*}\delta _{2*}
			\big)L^2
		\big]
		\nonumber
		\\[0.5em]
		&\qquad\quad
		+\,
		2\big(
			\epsilon _{1*}^2\delta _{1*}
			+
			\epsilon _{1*}\epsilon _{2*}\delta _{1*}
		\big)L
		-
		\big(
			\epsilon _{1*}\epsilon _{2*}\delta _{1*}
			+
			2\epsilon _{1*}\delta _{1*}\delta _{2*}
			-
			4\epsilon _{1*}\delta _{1*}^2
		\big)L^2
		\nonumber
		\\[0.5em]
		&\qquad\quad
		+\,
		\tfrac{1}{3}\,
		\big(
			4\delta _{1*}^3
			-
			6\delta _{1*}^2\delta _{2*}
			+
			\delta _{1*}\delta _{2*}^2
			+
			\delta _{1*}\delta _{2*}\delta _{3*}
		\big)L^3
		+
		\mathcal O\big((\epsilon _*,\delta _*)^4\big)
	\Big\}.
	\label{csSq}
\end{align}
Finally, for the index function $ \nu_s $ one can perform an expansion analogous to that for $ c_s^2 $ (see (\ref{csExpand})). Alternatively, noting that $ \nu_s ^2 \equiv \frac{1}{4} + \eta ^2U_s$, where the product $ \eta ^2 U_s $ by virtue of (\ref{etaThird}) and the first of (\ref{effU}) is expressible solely in terms of the Hubble and sound flow functions without the extra factor $(aH)$, we can simply use the expansions for these functions we have derived to write down the needed expression for $ \nu _s $. Both of these two methods give the same result:
\begin{align}
	\nu _s
	&=
	\tfrac{3}{2}
	+
	\big[
		\epsilon _{1*} + \tfrac{1}{2}\,\epsilon _{2*}
		+
		\delta _{1*}
	\big]
	+
	\big[
		\epsilon _{1*}^2
		+
		\epsilon _{1*}\delta _{1*}
		+
		\tfrac{11}{6}\,\epsilon _{1*}\epsilon _{2*}
		+
		\tfrac{1}{6}\,\epsilon _{2*}\epsilon _{3*}
		+
		\tfrac{1}{3}\,\delta _{1*}\delta _{2*}
		\nonumber
		\\[0.5em]
		&\qquad
		-\,
		\big(
			\epsilon _{1*}\epsilon _{2*}
			+
			\tfrac{1}{2}\,\epsilon _{2*}\epsilon _{3*}
			+
			\delta _{1*}\delta _{2*}
		\big)L
	\big]
	+
	\big[
		\epsilon _{1*}^3
		+
		\tfrac{77}{18}\,\epsilon _{1*}^2\epsilon _{2*}
		+
		\tfrac{17}{9}\,\epsilon _{1*}\epsilon _{2*}^2
		\nonumber
		\\[0.5em]
		&\qquad
		+\,
		\tfrac{14}{9}\,\epsilon _{1*}\epsilon _{2*}\epsilon _{3*}
		-
		\tfrac{1}{18}\,\epsilon _{2*}^2\epsilon _{3*}
		+
		\epsilon _{1*}^2\delta _{1*}
		+
		\tfrac{10}{9}\,\epsilon _{1*}\epsilon _{2*}\delta _{1*}
		-
		\tfrac{1}{9}\,\epsilon _{2*}\epsilon _{3*}\delta _{1*}
		\nonumber
		\\[0.5em]
		&\qquad
		+\,
		\tfrac{4}{9}\,\epsilon _{1*}\delta _{1*}\delta _{2*}
		-
		\tfrac{1}{9}\,\epsilon_{2*}\delta_{1*}\delta_{2*}
		-
		\tfrac{2}{9}\,\delta _{1*}^2\delta _{2*}
		-
		\big(
			3\epsilon _{1*}^2\epsilon _{2*}
			+
			\tfrac{11}{6}\,\epsilon _{1*}\epsilon _{2*}^2
			\nonumber
			\\[0.5em]
			&\qquad
			+\,
			\tfrac{7}{3}\,\epsilon _{1*}\epsilon _{2*}\epsilon _{3*}					+
			\tfrac{1}{6}\,\epsilon _{2*}\epsilon _{3*}^2
			+
			\tfrac{1}{6}\,\epsilon _{2*}\epsilon _{3*}\epsilon _{4*}
			+
			\epsilon _{1*}\epsilon _{2*}\delta _{1*}
			+
			2\epsilon _{1*}\delta _{1*}\delta _{2*}
			\nonumber
			\\[0.5em]
			&\qquad
			+\,
			\tfrac{1}{3}\,\delta _{1*}\delta _{2*}^2			
			+
			\tfrac{1}{3}\,\delta _{1*}\delta _{2*}\delta _{3*}
		\big)L
		+
		\big(
			\tfrac{1}{2}\,\epsilon _{1*}\epsilon _{2*}^2
			+
			\tfrac{1}{2}\,\epsilon _{1*}\epsilon _{2*}\epsilon _{3*}
			+
			\tfrac{1}{4}\,\epsilon _{2*}\epsilon _{3*}^2
			\nonumber
			\\[0.5em]
			&\qquad
			+\,
			\tfrac{1}{4}\,\epsilon _{2*}\epsilon _{3*}\epsilon _{4*}		
			+
			\tfrac{1}{2}\,\delta _{1*}\delta _{2*}^2
			+
			\tfrac{1}{2}\,\delta _{1*}\delta _{2*}\delta _{3*}
		\big)L^2
	\big]
	+
	\mathcal O\big((\epsilon _*,\delta _*)^4\big)
	\label{nuExpand}
\end{align}
Similarly, for the tensor perturbation index function we have
\begin{align}
	\nu _t
	&=
	\tfrac{3}{2}\,
	+
	\epsilon _{1*}
	+
	\big[
		\epsilon _{1*}^2 + \tfrac{4}{3}\,\epsilon _{1*}\epsilon _{2*}
		-
		\epsilon _{1*}\epsilon _{2*}L
	\big]
	+
	\big[
		\epsilon _{1*}^3
		+
		\tfrac{34}{9}\,\epsilon _{1*}^2\epsilon _{2*}
		+
		\tfrac{4}{3}\,\epsilon _{1*}\epsilon _{2*}^2		
		\nonumber
		\\[0.5em]
		&\qquad
		+\,
		\tfrac{4}{3}\,\epsilon _{1*}\epsilon _{2*}\epsilon _{3*}
		-
		\big(
			3\epsilon _{1*}^2\epsilon _{2*}
			+
			\tfrac{4}{3}\,\epsilon _{1*}\epsilon _{2*}^2
			+
			\tfrac{4}{3}\,\epsilon _{1*}\epsilon _{2*}\epsilon _{3*}
		\big)L
		\nonumber
		\\[0.5em]
		&\qquad
		+\,
		\big(
			\tfrac{1}{2}\,\epsilon _{1*}\epsilon _{2*}^2
			+
			\tfrac{1}{2}\,\epsilon _{1*}\epsilon _{2*}\epsilon _{3*}
		\big)L^2
	\big]
	+
	\mathcal O\big((\epsilon _*,\delta _*)^4\big)	
\end{align}
With these last two equations settled, we are now in a position to calculate the power spectra for scalar and tensor perturbations beyond the second order.

\bigskip
\bigskip
\section{Logarithmic Divergences in the Power Spectrum}
\label{logDiv}

We first consider the power spectrum for scalar perturbations given by (\ref{scalarPii}),
\begin{align}
	P_s 
	&=
	-\frac{k^3}{8\pi ^2}
	\frac{\eta c_s^2}{a^2 \epsilon _1 \nu _s}
	e^{2\Psi _s}.
	\label{powerSpec2}
\end{align}
Before we delve into the possible logarithmic divergences that may arise from this expression, noting that we have to take the limit $ \eta \rightarrow 0^- $, we address the possible problem coming from the factor $ \eta  $ in the right hand side. In fact, by virtue of the expansion for $ a^2 $ given by (\ref{aSq}) we have a total factor of $ \eta ^3 $ such that the zeroth-order power spectrum is
\begin{align}
	P_s^{(0)}
	=
	-\frac{k^3}{12\pi ^2}
	\frac{c_{s*}^2H_*^2}{\epsilon _{1*}}
	\eta ^3
	\big(e^{2\Psi _s}\big)^{(0)}.
\end{align}
A factor of $ 1/\eta ^3 $ should then fall out of the exponential term to avoid a trivial result. That this is so the case can be easily seen from the following calculation.
\begin{align}
	2\Psi _s
	&=
	2\int _{\eta _*}^\eta \dd\tau \,\sqrt{g_s}
	=
	-2\nu _*\int _{\eta _*}^\eta 
	\frac{\dd\tau }{\tau }\left(
		1 - \frac{k^2 c_{s*}^2\tau ^2}{\nu _{s*}^2}
	\right)^\frac{1}{2}
	+
	\cdots
	\nonumber
	\\[0.5em]
	2\Psi _s
	&=
	-3 + 3\ln 2 - 3L + \cdots.
	\label{firstTermDiv}
\end{align}
Exponentiation then yields
\begin{align}
	e^{2\Psi _s}
	&=
	8e^{-3}\left(\frac{\eta _*}{\eta }\right)^3\big(1 + \cdots\big),
\end{align}
which cancels $ \eta ^3 $ as advertised. 

We now go to the calculation of $ P_s $. With all the literal coefficients of $ e^{2\Psi _s} $ available from the previous section, all that is left is the evaluation of the exponential term. In order to do so, one expands the integrand about $(c_{s*}^2,\, \nu_{s*})$. 
\begin{align}
	g_s^\frac{1}{2}
	&=
	g_{s}^\frac{1}{2}\bigg|_{c_{s*}, \nu _{s*}}
	+\,
	\frac{\partial g^\frac{1}{2}}{\partial \nu_s }\bigg|_{\nu _{s*}, c_{s*}}
	\hspace{-1.5em}(\Delta \nu_s)
	\,+\,
	\frac{\partial g^\frac{1}{2}}{\partial c_s^2}\bigg|_{\nu _{s*}, c_{s*}}
	\hspace{-1.5em}
	(\Delta c_s^2) 
	+
	\cdots.
	\label{gsTaylor}
\end{align}
Here, $ \Delta c_s^2 = c_s^2 - c_{s*}^2 $ and $ \Delta \nu _s = \nu _s - \nu _{s*}$. Then $ \Psi _s $ can be rewritten as
\begin{align}
	\Psi _s
	&=
	-\nu _{s*}\int _1^w
	\frac{\dd u}{u}
	(1 - u^2)^\frac{1}{2}
	+
	\frac{\nu _{s*}}{2}
	\int_1^w
	\frac{\dd u\, u}{(1 - u^2)^\frac{1}{2}}(\Delta s)
	\nonumber
	\\[0.5em]
	&\qquad
	-\,
	\int _1^w \frac{\dd u}{u(1 - u^2)^\frac{1}{2}}(\Delta \nu _s)
	+
	\frac{\nu _{s*}}{8}\int _1^w \frac{\dd u\, u^3}{(1 - u^2)^\frac{3}{2}}
	(\Delta s)^2
	\nonumber
	\\[0.5em]
	&\qquad
	-\,
	\frac{1}{2}\int _1^w
	\frac{\dd u\,u}{(1 - u^2)^\frac{3}{2}}
	(\Delta s)(\Delta \nu _s)
	+
	\frac{\nu _{s*}}{16}\int _1^w \frac{\dd u\, u^5}{(1 - u)^\frac{5}{2}}
	(\Delta s)^3 + \cdots,
	\label{expandPsi}
\end{align}
where 
\begin{align}
	u \equiv -\frac{kc_{s*}\tau }{\nu _{s*}},
	\qquad
	w \equiv -\frac{kc_{s*}\eta}{\nu _{s*}},
	\quad
	\text{and}
	\quad
	\Delta s \equiv \frac{c_s^2 - c_{s*}^2}{c_{s*}^2}.
	\label{uws}
\end{align}
From (\ref{csSq}), $ (\Delta s) $ is at least first order while from (\ref{nuExpand}), $ (\Delta \nu _s) $ is at least second order. It follows that the remaining terms following the six integrals above are at least fourth order. Needless to say, the visible series of integrals in (\ref{expandPsi}) is enough to calculate $P_s$ up to third order. 

We further note that all the terms contained in $ \Delta s $ and $ \Delta \nu _s $ involve powers of logarithms; \textit{i.e.}, $ L^n = \ln ^n u$, where $ n \ge 1 $. As such, all the integrals above take the form
\begin{align}
	\int_1^w \dd u\, u^p\, \big(1 - u^2\big)^q\, \ln^r u,
\end{align}
where $ p,\, q, $ and $ r $ are all integers. This form allows us to determine the integrability of all the terms involved in the Taylor series expansion given by (\ref{gsTaylor}) and see where the logarithmically divergent terms may come from. The first term in this series expansion, by virtue of (\ref{firstTermDiv}), leads to a logarithmically divergent integral. For the remaining terms, we have the following classification : those involving (a) $(\Delta c_s^2)^n$ only, (b) $(\Delta\nu_s)^n$ only, and (c) $(\Delta\nu_s)^n(\Delta c_s^2)^p$, where $n,\, p \ge 1$.

The integrals involving only $ (\Delta c_s^2)^n $ or $(\Delta s)^n$ take the form
\begin{align}
	\int _1^w \dd u\,
	\frac{u^{2n-1}\,\ln^m u}{\big(1 - u^2\big)^{n - \frac{1}{2}}},
	\qquad
	\Big(
		m \ge n
	\Big)
\end{align}
The integrand is continuous on $ (0, 1) $. Furthermore, the limits as $ u\rightarrow 0^+ $ and $ u\rightarrow 1^- $ both exist and vanishing. It follows that all integrals involving $ (\Delta s)^n $ only are finite as $ w\rightarrow 0^+ $. Similar analysis indicates that all integrals involving the product $ (\Delta s)^n(\Delta \nu _s)^p $, where $ n, p \ge 1$, are also finite. The integrals involving only $ (\Delta \nu_s)^n $  take the form
\begin{align}
	\int _1^w
	\dd u \left[
	    \frac{c_1\ln ^m u}{u\big(1 - u^2\big)^{n - \frac{1}{2}}}
	    +
	    \frac{c_2\ln ^m u}{u\big(1 - u^2\big)^{n - \frac{3}{2}}}
	    +
	    \cdots
	\right],
	\qquad
	\Big(
		m \ge n
	\Big)
\end{align}
where $c_i$'s are constants. The number of terms inside the pair of square brackets depends on whether $n$ is odd or even. If it is odd, we have $(n + 1)/2$ terms while if it is even, we have $(n + 2)/2$. The integrand is continuous on $ (0, 1) $ and the limit as $ u\rightarrow 1^- $ is finite and vanishing. The limit however, of every term inside the pair of square brackets above as $ u \rightarrow 0^+ $ diverges. This leads to
\begin{align}
	\int _1^w
	\frac{\dd u\, \ln^m u}{u\big(1 - u^2\big)^{n - \frac{1}{2}}}
	=
	\frac{\ln^{m+1}w}{m + 1}
	+
	\text{const.}
\end{align}
in the limit as $ w\rightarrow 0^+ $. Hence, (a) all divergences involved in $ e^{2\Psi } $ are logarithmic in nature\footnote{Intermediate calculational results may involve polylogarithms of the form $ \text{Li}_s(\text{const}/w^n) $ which in the limit as $ w\rightarrow 0^+ $, result to logarithms.}, and (b) only those integrals involving $ (\Delta \nu _s)^n $ and \textit{not} $ (\Delta s)^n $ give rise to such logarithmic divergences.

Had we been only interested in logarithmic divergences in $ P_s $, we could have simply expressed $ c_s^2,\, a^2,\, \nu _s, $ and $ \epsilon _1 $ in exponential form and combine it with $ e^{2\Psi } $ so as to simply find whether we could have incomplete cancellation of terms involving $ \ln w $ without having to compute the finite integrals in (\ref{expandPsi}). Nonetheless, for completeness, we compute the full power spectrum at least up to third order. In order to do so, we perform the necessary integrations in (\ref{expandPsi}). Here are our results after a lengthy calculation.
\begin{align}
	A_0 ~&=~ 
	\int_{1}^{w}\dd u~
	\frac{\left(1-u^2\right)^\frac{1}{2}}{u}	
	=
	1 - \ln 2 + \ln w
	 \nonumber \\[0.5em]
	B_1 ~&=~ 	
	\int_{1}^{w}\dd u~
	\frac{u\;\ln u}{
		\left(1-u^2\right)^\frac{1}{2}
	}
	=
	1-\ln 2
	 \nonumber \\[0.5em]
	B_2 ~&=~ 	
	\int_{1}^{w}\dd u~
	\frac{u\;\ln^2 u}{
		\left(1-u^2\right)^\frac{1}{2}
	}
	=
	-2-\ln^2 2 + 2\ln 2 + \frac{\pi ^2}{12}
	 \nonumber \\[0.5em]
	B_3 ~&=~ 	
	\int_{1}^{w}\dd u~
	\frac{u\;\ln^3 u}{
		\left(1-u^2\right)^\frac{1}{2}
	}
	=		
	-\frac{3}{2}\zeta (3) - 6\ln 2 + 6 + \frac{\pi ^2}{4}\ln 2
	+
	3\ln ^2 2
	-
	\ln ^3 2
	-
	\frac{\pi ^2}{4}	
	 \nonumber \\[0.5em]
	C_1 ~&=~ 	
	\int_{1}^{w}\dd u~
	\frac{\ln u}{u\left(1-u^2\right)^\frac{1}{2}}	
	=		
	\frac{\pi ^2}{24} - \frac{1}{2}\ln^2 2 + \frac{1}{2}\ln^2 w
	 \nonumber \\[0.5em]
	C_2 ~&=~ 	
	\int_{1}^{w}\dd u~
	\frac{\ln^2 u}{u\left(1-u^2\right)^\frac{1}{2}}	
	=		
	-\frac{1}{2}\zeta (3) + \frac{\pi ^2}{12}\ln 2 - \frac{1}{3}\ln^3 2
	+
	\frac{1}{3}\ln^3 w
	 \nonumber \\[0.5em]
	D_2 ~&=~ 	
	\int_{1}^{w}\dd u~
	\frac{ u^3 \,\ln^2 u}{
		\left(1-u^2\right)^\frac{3}{2}
	}	
	=		
	2 + 2\ln^2 2 - 2\ln 2 - \frac{\pi ^2}{6}
	 \nonumber \\[0.5em]
	D_3 ~&=~ 	
	\int_{1}^{w}\dd u~
	\frac{ u^3 \,\ln^3 u}{
		\left(1-u^2\right)^\frac{3}{2}
	}	
	=		
	3\zeta (3) -6 - 3\ln^2 2 + 6\ln 2 - \frac{\pi ^2}{2}\ln 2
	+
	2\ln^3 2 + \frac{\pi ^2}{4}
	 \nonumber \\[0.5em]
	E_2 ~&=~ 	
	\int_{1}^{w}\dd u~
	\frac{u\,\ln^2 u}{
		\left(1-u^2\right)^\frac{3}{2}
	}	
	=		
	-\frac{\pi ^2}{12} + \ln^2 2
	 \nonumber \\[0.5em]
	F_3 ~&=~ 	
	\int_{1}^{w}\dd u~
	\frac{ u^5\,\ln^3 u}{
		\left(1-u^2\right)^\frac{5}{2}
	}	
	=		
	-4\zeta (3) + 6 + 2\ln^2 2 - \frac{8}{3}\ln^3 2
	+
	\frac{2\pi ^2}{3}\ln 2
	-
	6\ln 2 - \frac{\pi ^2}{6}
	\label{listInt}
\end{align}
The labels $ A,\,B,\,\cdots,F $ correspond respectively, to the visible six integrals in the right hand side of (\ref{expandPsi}). Moreover, the subscripts of these labels indicate the power of the log function in the integrand. 

Using the equations for definite integrals above in (\ref{expandPsi}) and then substituting the resulting equation together with the expansions for $ \epsilon _1 $, $ a^2 $, $ c_s^2 $, and $ \nu _s $ given by (\ref{eps1Expand}), (\ref{aSq}), (\ref{csSq}), and (\ref{nuExpand}) respectively, in the working equation for the power spectrum given by (\ref{powerSpec2}), and noting that $ kc_{s*} = -\nu _{s*}/\eta _* $ at the turning point, we find
\begin{align}
	P_s
	&=
	\frac{H_*^2 c_0}{8\pi ^2 \epsilon _{1*}c_{s*}}\big[
		\bar P_s^{(0)} + \bar P_s^{(1)} + \bar P_s^{(2)} + \bar P_s^{(3)}
	\big]
	\label{powerSpec3},
\end{align}
where $ c_0 = 18 e^{-3} \approx 0.896$ and
\begin{align}
	\bar P_s^{(0)}
	&=
	1
	\nonumber
	\\[1.0em]
	\bar P_s^{(1)}
	&=
	-\big(
		\tfrac{8}{3}
		-
		2 \ln 2
	\big)\epsilon_{1*}
	-
	\big(
		\tfrac{1}{3}
		-
		\ln 2		
	\big)\epsilon_{2*}
	+
	\big(
		\tfrac{7}{3}\,
		-
		 \ln 2
	\big)\delta_{1*}
	\nonumber
	\\[1.0em]
	\bar P^{(2)}_s
	&=
	\big(
		\tfrac{13}{9}
		-
		\tfrac{10}{3}\, \ln 2 
		+
		2 \ln^2 2
	\big)\epsilon_{1*}^2
	-
	\big(
		\tfrac{25}{9}
		-	
		\tfrac{1}{12}\, \pi^2 
		-
		\tfrac{1}{3}\, \ln 2 		
		-
		\ln^2 2
	\big)\epsilon_{1*} \epsilon_{2*}
	\nonumber
	\\[0.5em]
	&\qquad
	-\,
	\big(
		\tfrac{1}{18}
		+
		\tfrac{1}{3}\, \ln 2		
		-	
		\tfrac{1}{2}\, \ln^2 2
	\big) \epsilon_{2*}^2
	-
	\big(
		\tfrac{1}{9}
		-	
		\tfrac{1}{24}\, \pi^2
		-
		\tfrac{1}{3}\, \ln 2 		
		+
		\tfrac{1}{2}\, \ln^2 2
	\big)\epsilon_{2*} \epsilon_{3*}
	\nonumber
	\\[0.5em]
	&\qquad
	-\,
	\big(
		\tfrac{25}{9}
		-	
		\tfrac{13}{3}\, \ln 2 
		+
		2 \ln^2 2
	\big)\epsilon_{1*} \delta_{1*}
	-
	\big(
		\tfrac{2}{9}
		-
		\tfrac{5}{3}\, \ln 2 
		+
		\ln^2 2		
	\big)\epsilon_{2*} \delta_{1*}
	\nonumber
	\\[0.5em]
	&\qquad
	+\,
	\big(
		\tfrac{23}{18}
		-
		\tfrac{4}{3}\, \ln 2 		
		+
		\tfrac{1}{2}\, \ln^2 2	
	\big) \delta_{1*}^2 
	+
	\big(
		\tfrac{25}{9}	
		-
		\tfrac{1}{24}\, \pi^2		
		-
		\tfrac{7}{3}\, \ln 2 
		+
		\tfrac{1}{2}\, \ln^2 2
	\big)\delta_{1*} \delta_{2*}
	\nonumber
	\\[1.0em]
	\bar P^{(3)}_s
	&=
	\big(
		\tfrac{4}{9}
		-
		\tfrac{4}{9}\, \ln 2		
		-
		\tfrac{4}{3}\, \ln^2 2    
		+
		\tfrac{4}{3}\, \ln^3 2 
	\big)\epsilon_{1*}^3
	+
	\big(
		\tfrac{2}{3}  
		+
		\tfrac{1}{36}\, \pi^2		
		+
		\tfrac{1}{6}\, \pi^2 \ln 2 
		-
		6 \ln 2  
		+
		3 \ln^2 2    
	\big)\epsilon_{1*}^2 \epsilon_{2*}
	\nonumber
	\\[0.5em]
	&\qquad
	-\,
	\big(
		\tfrac{23}{9}\, 
		-
		\tfrac{1}{8}\, \pi^2
		-
		\tfrac{1}{2}\, \zeta(3)				
		+
		\tfrac{1}{3}\, \ln 2    
		-
		\tfrac{1}{6}\, \ln^2 2  
		-
		\tfrac{1}{3}\, \ln^3 2    
	\big)\epsilon_{1*} \epsilon_{2*}^2
	\nonumber
	\\[0.5em]
	&\qquad
	+\,
	\big(
		\tfrac{1}{18}
		-
		\tfrac{1}{18}\, \ln 2 		  
		-
		\tfrac{1}{6}\, \ln^2 2    
		+	
		\tfrac{1}{6}\, \ln^3 2		
	\big)\epsilon_{2*}^3
	\nonumber
	\\[0.5em]
	&\qquad
	-\,
	\big(
		\tfrac{26}{9}
		-
		\tfrac{1}{12}\, \pi^2		
		-
		\tfrac{1}{2}\, \zeta(3)		   
		-
		2 \ln 2   
		+
		\tfrac{1}{3}\, \ln^2 2  
		+	
		\tfrac{2}{3}\, \ln^3 2  		
	\big)\epsilon_{1*} \epsilon_{2*} \epsilon_{3*}
	\nonumber
	\\[0.5em]
	&\qquad
	-\,
	\big(
		\tfrac{1}{72}\, \pi^2 
		-	
		\tfrac{1}{24}\, \pi^2 \ln 2 
		+
		\tfrac{1}{3}\, \ln 2  
		+
		\tfrac{1}{2}\, \ln^3 2 
		-
		\tfrac{1}{2}\, \ln^2 2 
	\big)\epsilon_{2*}^2 \epsilon_{3*}
	\nonumber
	\\[0.5em]
	&\qquad
	+\,
	\big(
		\tfrac{1}{72}\, \pi^2  
		+	
		\tfrac{1}{4}\, \zeta(3)	
		-
		\tfrac{1}{24}\, \pi^2 \ln 2  
		-
		\tfrac{1}{6}\, \ln^2 2    
		+
		\tfrac{1}{6}\, \ln^3 2  
	\big)\epsilon_{2*} \epsilon_{3*}^2
	\nonumber
	\\[0.5em]
	&\qquad
	+\,
	\big(
		\tfrac{1}{72}\, \pi^2  
		+
		\tfrac{1}{4}\, \zeta(3) 
		-
		\tfrac{1}{24}\, \pi^2 \ln 2  		 		
		-
		\tfrac{1}{6}\, \ln^2 2  
		+
		\tfrac{1}{6}\, \ln^3 2  
	\big)\epsilon_{2*} \epsilon_{3*} \epsilon_{4*}
	\nonumber
	\\[0.5em]
	&\qquad
	-\,
	\big(
		\tfrac{2}{3}  
		-
		2 \ln^2 2 
		-
		\tfrac{2}{3}\, \ln 2  
		+
		2 \ln^3 2 
	\big)\epsilon_{1*}^2 \delta_{1*}
	\nonumber
	\\[0.5em]
	&\qquad
	-\,
	\big(
		\tfrac{2}{3} 
		+
		\tfrac{1}{72}\, \pi^2		     
		-
		\tfrac{10}{3}\, \ln 2 
		+
		\tfrac{1}{12}\, \pi^2 \ln 2 
		+
		\tfrac{1}{2}\, \ln^2 2
		+
		\ln^3 2				
	\big)\epsilon_{1*} \epsilon_{2*} \delta_{1*}
	\nonumber
	\\[0.5em]
	&\qquad
	-\,
	\big(
		\tfrac{1}{6}
		-
		\tfrac{1}{6}\, \ln 2 
		-	
		\tfrac{1}{2}\, \ln^2 2 		
		+
		\tfrac{1}{2}\, \ln^3 2 
	\big)\epsilon_{2*}^2 \delta_{1*}
	\nonumber
	\\[0.5em]
	&\qquad
	+\,
	\big(
		\ln^3 2 
		-
		\ln^2 2 
		+
		\tfrac{1}{3}\,  
		-
		\tfrac{1}{3}\, \ln 2  
	\big)\epsilon_{1*} \delta_{1*}^2
	+
	\big(
		\tfrac{1}{6} 	
		-
		\tfrac{1}{6}\, \ln 2  
		-
		\tfrac{1}{2}\, \ln^2 2 
		+
		\tfrac{1}{2}\, \ln^3 2 
	\big)\epsilon_{2*} \delta_{1*}^2
	\nonumber
	\\[0.5em]
	&\qquad
	+\,
	\big(
		\tfrac{1}{72}\, \pi^2
		+	
		\tfrac{1}{3}\, \ln 2  
		-
		\tfrac{1}{24}\, \pi^2 \ln 2  		
		-
		\tfrac{1}{2}\, \ln^2 2  
		+
		\tfrac{1}{2}\, \ln^3 2 		
	\big)\epsilon_{2*} \epsilon_{3*} \delta_{1*}
	\nonumber
	\\[0.5em]
	&\qquad
	-\,
	\big(
		\tfrac{1}{18}	
		-
		\tfrac{1}{18}\, \ln 2  		
		-
		\tfrac{1}{6}\, \ln^2 2  
		+
		\tfrac{1}{6}\, \ln^3 2 		
	\big)\delta_{1*}^3
	\nonumber
	\\[0.5em]
	&\qquad
	-\,
	\big(
		\tfrac{1}{18}\, \pi^2 
		-	
		\tfrac{14}{3}\, \ln 2   
		+
		\tfrac{1}{12}\, \pi^2 \ln 2  
		+
		4 \ln^2 2   		
		-
		\ln^3 2  
	\big)\epsilon_{1*} \delta_{1*} \delta_{2*}
	\nonumber
	\\[0.5em]
	&\qquad
	-\,
	\big(
		\tfrac{1}{36}\, \pi^2 
		-	
		\tfrac{7}{3}\, \ln 2  
		+
		\tfrac{1}{24}\, \pi^2 \ln 2  
		+
		2 \ln^2 2  		
		-
		\tfrac{1}{2}\, \ln^3 2  
	\big)\epsilon_{2*} \delta_{1*} \delta_{2*}
	\nonumber
	\\[0.5em]
	&\qquad
	+\,
	\big(
		\tfrac{1}{36}\, \pi^2 	
		-
		\tfrac{7}{3}\, \ln 2  
		+
		\tfrac{1}{24}\, \pi^2 \ln 2 
		+
		2  \ln^2 2		
		-
		\tfrac{1}{2}\, \ln^3 2  
	\big)\delta_{1*}^2 \delta_{2*} 
	\nonumber
	\\[0.5em]
	&\qquad
	+\,
	\big(
		3 	
		-
		\tfrac{7}{72}\, \pi^2  
		-
		\tfrac{1}{4}\, \zeta(3)  		
		-
		3 \ln 2  				
		+
		\tfrac{1}{24}\, \pi^2 \ln 2 			
		+	
		\tfrac{7}{6}\, \ln^2 2   
		-
		\tfrac{1}{6}\, \ln^3 2  
	\big)\delta_{1*} \delta_{2*}^2
	\nonumber
	\\[0.5em]
	&\qquad
	+\,
	\big(
		3	
		-
		\tfrac{7}{72}\, \pi^2 		
		-
		\tfrac{1}{4}\,  \zeta(3) 		
		-
		3 \ln 2 		
		+
		\tfrac{1}{24}\, \pi^2 \ln 2  		
		+
		\tfrac{7}{6}\,  \ln^2 2 
		-
		\tfrac{1}{6}\, \ln^3 2   
	\big) \delta_{1*} \delta_{2*} \delta_{3*}
	\nonumber
	\\[0.5em]
	&\qquad
	+\,
	\tfrac{1}{9}\big(
		2 \epsilon_{1*} \epsilon_{2*}^2 
		+
		2 \epsilon_{1*} \epsilon_{2*} \epsilon_{3*}
		+
		\epsilon_{2*} \epsilon_{3*}^2
		+
		\epsilon_{2*} \epsilon_{3*} \epsilon_{4*}
		+
		2\delta_{1*} \delta_{2*}^2 
		+	
		2\delta_{1*} \delta_{2*} \delta_{3*} 
	\big)\ln w.
	\label{psThree}
\end{align}
The result above for the power spectrum generalises the second-order approximation for $ P_s $ given in Ref. \cite{Martin:2013}. The third-order part of (\ref{powerSpec3}) is something new. What is surprising about this part (see the expression for $ \bar P^{(3)}_s $), is the unexpected set of terms involving $ \ln w $ which is divergent in the limit as $ w\rightarrow 0^+ $. Unlike that of $ P_s^{(0)} $, $ P_s^{(1)} $, and $ P_s^{(2)} $, we have an incomplete cancellation of (divergent) log terms. Furthermore, considering the case of a constant speed of sound does not help eliminate these terms. Since terms involving $ \ln w$ are \textit{carried} to the calculation of the higher-order power spectrum, we expect $ \bar P^{(n)}_s $ for $ n \ge 4 $ to also contain $ \ln w $ (raised to some positive integer).

The calculation for the power spectrum for the tensor perturbations is less cumbersome than that of the scalar perturbations. The speed of sound is unity and out of the 10 integrals used for $ P_s $, we only need three corresponding to the first and third terms in the right hand side of (\ref{expandPsi}). Following a similar approach, now with (\ref{tensorP}) as the working equation for $ P_t $ instead of (\ref{powerSpec2}) for $ P_s $, we find
\begin{align}
	P_t
	&=
	\frac{2H_*^2c_0}{\pi ^2}\Big[
		\bar P_t^{(0)} + \bar P_t^{(1)} + \bar P_t^{(2)} + \bar P_t^{(3)}
	\Big],
\end{align}
where
\begin{align}
	\bar P_t^{(0)}
	&=
	1
	\nonumber
	\\[1.0em]
	\bar P_t^{(1)}
	&=
	\big(-\tfrac{8}{3}
	+
	2\ln 2\big)\epsilon _{1*}
	\nonumber
	\\[1.0em]
	\bar P^{(2)}_t
	&=
	\big(
		\tfrac{13}{9}
		-
		\tfrac{10}{3}\,\ln 2
		+
		2\ln^2 2
	\big)\epsilon _{1*}^2
	-
	\big(
		\tfrac{26}{9}
		-
		\tfrac{1}{12}\,\pi ^2
		-
		\tfrac{8}{3}\,\ln 2
		+
		\ln^2 2
	\big)\epsilon _{1*}\epsilon _{2*}
	\nonumber
	\\[1.0em]
	\bar P^{(3)}_t
	&=
	\big(
		\tfrac{4}{9}
		-
		\tfrac{4}{9}\, \ln 2  
		-
		\tfrac{4}{3}\, \ln^2 2 
		+
		\tfrac{4}{3}\, \ln^3 2 
	\big)\epsilon_{1*}^3 
	\nonumber
	\\[0.5em]
	&\qquad
	+\,
	\big(
		\tfrac{1}{36}\, \pi^2 
		-
		\tfrac{16}{3}\,  \ln 2 
		+
		\tfrac{1}{6}\, \pi^2 \ln 2 
		+
		5 \ln^2 2 
		-
		2 \ln^3 2 
	\big)\epsilon_{1*}^2 \epsilon_{2*} 
	\nonumber
	\\[0.5em]
	&\qquad
	-\,
	\big(
		\tfrac{26}{9}
		-
		\tfrac{1}{9}\, \pi^2 
		-
		\tfrac{1}{2}\, \zeta(3) 
		-
		\tfrac{8}{3}\, \ln 2  
		+
		\tfrac{1}{12}\, \pi^2 \ln 2 
		+
		\tfrac{4}{3}\, \ln^2 2  
		-
		\tfrac{1}{3}\, \ln^3 2 
	\big)\epsilon_{1*} \epsilon_{2*}^2 
	\nonumber
	\\[0.5em]
	&\qquad
	-\,
	\big(
		\tfrac{26}{9}
		-
		\tfrac{1}{9}\, \pi^2 
		-
		\tfrac{1}{2}\, \zeta(3) 
		-
		\tfrac{8}{3}\, \ln 2  
		+
		\tfrac{1}{12}\, \pi^2 \ln 2 
		+
		\tfrac{4}{3}\, \ln^2 2 
		-
		\tfrac{1}{3}\, \ln^3 2 
	\big)\epsilon_{1*} \epsilon_{2*} \epsilon_{3*} 
	\nonumber
	\\[0.5em]
	&\qquad
	+\,
	\tfrac{2}{9}\big(
		\epsilon_{1*} \epsilon_{2*}^2 
		+
		\epsilon_{1*} \epsilon_{2*} \epsilon_{3*}
	\big) \ln w 
\end{align}
Just like that of the power spectrum for scalar perturbations we have an incomplete cancellation of (divergent) log terms. 

Let us try to look further into the details of the emergence of these logarithmically divergent terms in the power spectrum for the scalar perturbations, if only to convince us of such incomplete cancellation. (A similar analysis holds for tensor perturbations). In (\ref{psThree}) there are six terms involving $\ln w$. 
\begin{align}
    \bar P^{(3)}_s
    &\supset
	\big(
		\tfrac{2}{9} \epsilon_{1*} \epsilon_{2*}^2 
		+
		\tfrac{2}{9} \epsilon_{1*} \epsilon_{2*} \epsilon_{3*}
		+
		\tfrac{1}{9}\epsilon_{2*} \epsilon_{3*}^2
		+
		\tfrac{1}{9}\epsilon_{2*} \epsilon_{3*} \epsilon_{4*}
		+
		\tfrac{2}{9}\delta_{1*} \delta_{2*}^2 
		+	
		\tfrac{2}{9}\delta_{1*} \delta_{2*} \delta_{3*} 
	\big)\ln w
	\label{p3Log}
\end{align}
Terms in the working equation for $P_s$ contributing to this set of log terms may be seen by rewriting (\ref{powerSpec2}) in exponential form as alluded to earlier. To do this, use the definitions of the Hubble and sound flow functions for $\epsilon_1$ and $c_s^2$. For $a^2$ note that $\dd a/a = -(\dd\eta/\eta)(1 + \alpha)$ where the conformal time is written as $\eta = -(1 + \alpha)/(aH)$. For $\nu_s$ observe that  $\Delta \nu_s = \nu_s - \nu_{s*}$ is at least second order so $(\Delta \nu_s)^2$ can be truncated in the exponentiation. With these considerations, we can rewrite $\epsilon_1,\, c_s^2,\, a^2,$ and $\nu_s$ in exponential form as
\begin{align}
    \epsilon _1
	&=
	\epsilon _{1*}\exp\left[-\int_{\eta_*}^\eta \frac{\dd \tau }{\tau }\,
	\epsilon _2(1 + \alpha )\right],
	\nonumber
    \\[0.5em]
    c_s^2
	&=
	c_{s*}^2\exp\left[-2\int_{\eta_*}^\eta 
	\frac{\dd \tau }{\tau }\delta _1(1 + \alpha )\right],
	\nonumber
    \\[0.5em]
    a^2
	&=
	\frac{(1 + \alpha _*)^2}{(H_*\eta )^2}
	\exp\left(-2\int_{\eta_*}^\eta \frac{\dd \tau }{\tau }\alpha \right),
	\nonumber
    \\[0.5em]
    \frac{1}{\nu }
	&=
	\frac{1}{\nu _*} e^{-\frac{\Delta \nu }{\nu _*}}
\end{align}
where the last equation only holds up to third order. Substituting into the working equation for $P_s$ given by (\ref{powerSpec2})  we find
\begin{align}
    P_s
    &=
	\mathcal N(\eta)\exp\bigg\{
		\int_{\eta_*}^\eta \frac{\dd \tau }{\tau }\Big[
			\alpha (2 + \epsilon _2 + 2\delta _1)
			+
			\epsilon _2 + 2\delta _1
		\Big]
		-
		\frac{\Delta \nu }{\nu _*}
		+
		2\Psi_s 
	\bigg\},
	\label{powerSpecRewrite}
\end{align}
where 
\begin{align}
    \mathcal N
    &\equiv
    -\frac{k^3}{4\pi^2}
    \frac{H_*^2 c_{s*}^2}{\epsilon_{1*}\nu_{s*}(1 + \alpha_*)^2}
    \eta^3.
\end{align}

We can now focus on the three-term argument of the exponential function in (\ref{powerSpecRewrite}). Specifically, noting that all third-order terms involving $\ln^3 w$ and $\ln^2 w$ are canceled along the way, we pay attention to those third-order terms involving only the first power of $\ln w$. Considering $2\Psi_s$, a third-order term involving the first power of $\ln w$ can only come from the first integral in the right hand side of (\ref{expandPsi}) since all integrals involving $(\Delta s)^n$ are convergent and the divergent integrals involving $(\Delta\nu_s)^n$ start at $\ln^2 w$ (see the fifth and sixth of (\ref{listInt})). It follows that $2\Psi_s$ only contributes the following third-order logarithmically divergent terms involving the first power of $\ln w$:
\begin{align}
    2\Psi_s
    &\supset
    -\Big(
		2\epsilon _{1*}^3
		+
		\tfrac{28}{9}\,\epsilon _{1*}\epsilon _{2*}\epsilon _{3*}
		+
		\tfrac{34}{9}\,\epsilon _{1*}\epsilon _{2*}^2
		+
		\tfrac{77}{9}\,\epsilon _{1*}^2\epsilon _{2*}
		\nonumber
		\\[0.5em]
		&\qquad
		-\,
		\tfrac{1}{9}\,\epsilon _{2*}^2\epsilon _{3*}
		+
		2\epsilon _{1*}^2\delta _{1*} 		
		+
		\tfrac{8}{9}\,\epsilon _{1*}\delta _{1*}\delta _{2*}
		+
		\tfrac{20}{9}\,\epsilon _{1*}\epsilon _{2*}\delta _{1*}
		\nonumber
		\\[0.5em]
		&\qquad	
		-
		\tfrac{2}{9}\,\epsilon _{2*}\delta _{1*}\delta _{2*}
		-
		\tfrac{2}{9}\,\epsilon _{2*}\epsilon _{3*}\delta _{1*}
		-
		\tfrac{4}{9}\,\delta _{1*}^2\delta _{2*}
	\Big)\ln w
\end{align}
The second term in the argument of the exponential function in (\ref{powerSpecRewrite}) contributes
\begin{align}
    -\frac{\Delta\nu_s}{\nu_{s*}}
    &\supset
    \Big(
		\tfrac{14}{9}\,\epsilon _{1*}^2\epsilon _{2*} 
		+
		\epsilon_{1*}\epsilon _{2*}^2
		+
		\tfrac{4}{3}\,\epsilon _{1*}\epsilon _{2*}\epsilon _{3*}
		-
		\tfrac{1}{9}\,\epsilon _{2*}^2\epsilon _{3*}
		+
		\tfrac{2}{9}\,\epsilon _{1*}\epsilon _{2*}\delta _{1*}		
		\nonumber		
		\\[0.5em]
		&\qquad
		+\,
		\tfrac{8}{9}\,\epsilon _{1*}\delta _{1*}\delta _{2*}
		-
		\tfrac{2}{9}\,\epsilon _{2*}\epsilon _{3*}\delta _{1*}
		-
		\tfrac{2}{9}\,\epsilon _{2*}\delta _{1*}\delta _{2*}
		-
		\tfrac{4}{9}\,\delta _{1*}^2\delta _{2*}		
		\nonumber		
		\\[0.5em]
		&\qquad
		\underline{
    		+\,
    		\tfrac{1}{9}\,\epsilon _{2*}\epsilon _{3*}^2
    		+
    		\tfrac{1}{9}\,\epsilon _{2*}\epsilon _{3*}\epsilon _{4*}		
    		+\,
    		\tfrac{2}{9}\,\delta _{1*}\delta _{2*}\delta _{3*}		
    		+
    		\tfrac{2}{9}\,\delta _{1*}\delta _{2*}^2
    	}
	\Big)\ln w
\end{align}
Already at this point, one sees the last four terms above inside the pair of parentheses match the last four terms inside the pair of parentheses in the expression for $\bar P_{s}^{(3)}$ given by (\ref{p3Log}). These four terms are nowhere to be found in the contribution of $2\Psi_s$. Furthermore, writing out the integrand in (\ref{powerSpecRewrite}) as
\begin{align}
    \alpha (2 + \epsilon _2 + 2\delta _1) + \epsilon _2 + 2\delta _1
	&=
	(2\epsilon _1 + \epsilon _2 + 2\delta _1)
	+
	(3\epsilon _1\epsilon _2 + 2\epsilon _1^2 + 2\epsilon _1 \delta _1)
	\nonumber
	\\[0.5em]
	&\qquad
	+\,
	(
    	2\epsilon _1\epsilon _2\epsilon _3 + 3\epsilon _1\epsilon _2^2
    	+
    	7\epsilon _1^2\epsilon _2 + 2\epsilon _1^3
    	+
    	2\epsilon _1\epsilon _2\delta _1
    	+
    	2\epsilon _1^2\delta _1
	),	
	\label{expandAlpha}
\end{align}
tells us that with the current expansion scheme, there is no way to produce the (negative of the) mentioned four terms out of the integration of the last equation above. For instance, $\delta_{1*} \delta_{2*} \delta_{3*}\ln w$ can only possibly be produced from the third term $(2\delta_1)$ in the right hand side of (\ref{expandAlpha}). But the expansion of $2\delta_1$ only leads to the closest possible term $\delta_{1*} \delta_{2*} \delta_{3*}\ln^2 w$ which upon integration yields $\frac{1}{3}\,\delta_{1*} \delta_{2*} \delta_{3*}\ln^3 w$. Upon further considering the remaining three terms namely, $\tfrac{1}{9}\,\epsilon _{2*}\epsilon _{3*}^2,$ $\tfrac{1}{9}\,\epsilon _{2*}\epsilon _{3*}\epsilon _{4*},$ and $\tfrac{2}{9}\,\delta _{1*}\delta _{2*}^2$, we find that the second argument of the exponential function in (\ref{powerSpecRewrite}) namely, $-(\Delta\nu_s)/\nu_s$, is the sole contributor of four out of six logarithmically divergent terms in $\bar P^{(3)}_s$. The remaining two are a product of incomplete cancellation from the three-term argument of the exponential function in (\ref{powerSpecRewrite}).

Hence, with the expansion scheme laid down in Ref. \cite{Martin:2013} (thoroughly reviewed and elaborated in Sec. \ref{seriesExpandOther}) and the method of the UA to leading order of precision, the power spectra for scalar and tensor perturbations fail to converge starting at the third order with respect to the Hubble and sound flow parameters. Looking back, we see that the expansion scheme already involves divergent terms in the limit as $ \eta \rightarrow 0^- $. For instance, 
\begin{align}
	\epsilon _1
	&=
	\epsilon_{1*} - \epsilon _{1*}\epsilon _{2*}\ln \frac{\eta }{\eta _*}
	+
	\cdots.
\end{align}
One then delays taking the limit to the end of the calculation of the power spectrum in the hope that all the possibly logarithmically divergent terms would cancel along the way. It turns out that this technique only works up to second order. 

We wish to emphasise that the failure to converge cannot be attributed to the UA itself. The working equations for the power spectra given by (\ref{scalarPii}) and (\ref{tensorP})---the main machinery of the UA in calculating $ P_s $ and $ P_t $--- are  mathematically well-defined (finite) expressions. This leads us to put the blame on the implementation of the working equation itself. When applied to the case of the slow-roll \textit{k}-inflationary scenario to obtain quantities appropriately decomposed with respect to the Hubble and sound flow parameters,  this implementation requires a suitable expansion scheme for the slow-roll parameter, speed of sound, etc. As we have found, this scheme has brought with itself terms that can blow up in the limit as $ \eta \rightarrow 0^- $, that eventually fail to completely cancel in the end result starting at the third order for the power spectra. To be clear, the origin of the divergences is the expansion scheme itself, not the UA.

It is worth noting that an ``alternative'' expansion scheme was considered in Refs. \cite{Habib:2005mh,Zhu:2014aea}. It involves series expansion about the turning point but the variable of differentiation is the conformal time instead of the number of $e$-folds in the current scheme. For instance,
\begin{align}
	\label{altExpand}
    \nu_s
    &=
    \nu_{s*}
    +
    \frac{\dd\nu_s}{\dd\eta}\bigg|_{*}
    (\eta - \eta_*)
    +
    \frac{1}{2!}\frac{\dd^2\nu_s}{\dd\eta^2}\bigg|_{*}
    (\eta - \eta_*)^2
    +
    \cdots
\end{align}
However, (although the proof requires a lengthy calculation,) this expansion scheme is equivalent to that covered in Subsec. \ref{seriesExpandOther}; therefore, it is not an ``alternative'' one. Furthermore, using this in its current form is practically cumbersome at least for the inflation model we are considering here. In agreement with Ref. \cite{Zhu:2014aea}, the higher-order derivative terms for slow-roll inflation can contain lower order terms with respect to the Hubble and sound flow parameters. It follows that ignoring higher-order derivative terms amounts to a significant loss of precision. References \cite{Zhu:2014aea} and \cite{Zhu:2014wfa} adopted the current expansion scheme (discussed in Subsec. \ref{seriesExpandOther}) to avoid this problem in their calculations of the power spectra up to second order in $(\epsilon_{n*}, \delta_{n*})$ and up to third-order precision in the UA. 

\bigskip
\bigskip
\section{Tensor-to-scalar Ratio, Spectral Index, and Running}
\label{specRunTensor}

\bigskip
\subsection{Tensor-to-scalar Ratio}
The tensor-to-scalar ratio is defined as $ r \equiv P_t/P_s $. Since the logarithmically divergent term $ \ln w $ starts to appear in the third-order part of both (approximations for) $ P_s $ and $ P_t $, one may expect that the third-order part of $ r $ should contain $ \ln w $ as well. That this is not the case can be seen by substituting the expressions for $ P_s $ and $ P_t $ given by (\ref{scalarPii}) and (\ref{tensorP}) in the definition of $ r $ above.
\begin{align}
	r
	&=
	16\frac{\epsilon _1 }{c_s^2}
	\frac{\nu _s}{\nu _t}
	e^{2(\Psi _t - \Psi _s)}.
\end{align}
Here, the limit as $ \eta \rightarrow 0^- $ is again implicitly assumed. The term $ \epsilon _1 $ in the denominator in the expression for $ P_s $ is brought up making the zeroth-order part of $ r $ vanishing. As a consequence, the tensor-to-scalar ratio can be computed up to third order without encountering logarithmic divergence. Here is our result.
\begin{align}
	r
	&=
	16\epsilon _{1*}c_{s*}\big[
		\bar r^{(0)} + \bar r^{(1)} + \bar r^{(2)} + \bar r^{(3)}
	\big],
\end{align}
where
\begin{align}
	\bar r^{(0)}
	&=
	1
	\nonumber
	\\[1.0em]
	\bar r^{(1)}
	&=
	\big(
		\tfrac{1}{3} - \ln 2
	\big)\epsilon _{2*}
	-
	\big(
		\tfrac{7}{3}\, - \ln 2
	\big)\delta _{1*}
	\nonumber
	\\[1.0em]
	\bar r^{(2)}	
	&=
	\big(
		\tfrac{7}{9}
		-
		\ln 2 
	\big)\epsilon_{1*} \epsilon_{2*} 
		+
	\big(
		\tfrac{1}{6}
		-
		\tfrac{1}{3}\, \ln 2 
		+
		\tfrac{1}{2}\, \ln^2 2 
	\big) \epsilon_{2*}^2
	+
	\big(
		\tfrac{1}{9}
		-
		\tfrac{1}{24}\, \pi^2  
		-
		\tfrac{1}{3}\, \ln 2 
		+
		\tfrac{1}{2}\, \ln^2 2 
	\big)\epsilon_{2*} \epsilon_{3*}
	\nonumber
	\\[0.5em]
	&\qquad
	-\,
	\big(
		\tfrac{31}{9} 
		-
		3 \ln 2 
	\big)\epsilon_{1*} \delta_{1*} 
	-
	\big(
		\tfrac{4}{3}
		-
		\tfrac{11}{3}\, \ln 2 
		+
		\ln^2 2 
	\big)\epsilon_{2*} \delta_{1*} 
	+
	\big(
		\tfrac{25}{6}
		-
		\tfrac{10}{3}\, \ln 2 
		+
		\tfrac{1}{2}\, \ln^2 2 
	\big)\delta_{1*}^2 
	\nonumber
	\\[0.5em]
	&\qquad
	-\,
	\big(
		\tfrac{25}{9}
		-
		\tfrac{1}{24}\,  \pi^2 
		-
		\tfrac{7}{3}\, \ln 2 
		+
		\tfrac{1}{2}\, \ln^2 2 
	\big)\delta_{1*} \delta_{2*}
	\nonumber
	\\[1.0em]
	\bar r^{(3)}
	&=
	\big(
		\tfrac{25}{27}\, 
		-
		\ln 2 
	\big)\epsilon_{1*}^2 \epsilon_{2*} 
	+
	\big(
		\tfrac{35}{27}
		-
		\tfrac{1}{24}\, \pi^2  
		-
		\tfrac{20}{9}\,\ln 2   
		+
		\tfrac{3}{2}\, \ln^2 2  
	\big)\epsilon_{1*} \epsilon_{2*}^2 
	\nonumber
	\\[0.5em]
	&\qquad
	+\,
	\big(
		\tfrac{1}{54} 
		-
		\tfrac{1}{6}\, \ln 2 
		+
		\tfrac{1}{6}\,  \ln^2 2 
		-
		\tfrac{1}{6}\, \ln^3 2  
	\big)\epsilon_{2*}^3
	+
	\big(
		\tfrac{8}{27}
		-
		\tfrac{1}{12}\, \pi^2  
		-
		\tfrac{4}{9}\, \ln 2 
		+
		\ln^2 2  
	\big)\epsilon_{1*} \epsilon_{2*} \epsilon_{3*} 
	\nonumber
	\\[0.5em]
	&\qquad
	+\,
	\big(
		\tfrac{2}{27}
		-
		\tfrac{1}{72}\, \pi^2 
		-
		\tfrac{1}{9}\, \ln 2  
		+
		\tfrac{1}{24}\, \pi^2 \ln 2 
		+
		\tfrac{1}{2}\, \ln^2 2 
		-
		\tfrac{1}{2}\, \ln^3 2 
	\big)\epsilon_{2*}^2 \epsilon_{3*} 
	\nonumber
	\\[0.5em]
	&\qquad
	-\,
	\big(
		\tfrac{1}{72}\, \pi^2 
		+
		\tfrac{1}{4}\, \zeta(3)  
		-
		\tfrac{1}{24}\, \pi^2 \ln 2  
		-
		\tfrac{1}{6}\, \ln^2 2  
		+
		\tfrac{1}{6}\,  \ln^3 2 
	\big)\epsilon_{2*} \epsilon_{3*}^2 
	\nonumber
	\\[0.5em]
	&\qquad
	-\,
	\big(
		\tfrac{1}{72}\, \pi^2 
		+
		\tfrac{1}{4}\, \zeta(3) 
		-
		\tfrac{1}{24}\, \pi^2 \ln 2 
		-
		\tfrac{1}{6}\,  \ln^2 2 
		+
		\tfrac{1}{6}\, \ln^3 2 
	\big)\epsilon_{2*} \epsilon_{3*} \epsilon_{4*} 
	\nonumber
	\\[0.5em]
	&\qquad
	-\,
	\big(
		\tfrac{139}{27}\,  
		-
		5 \ln 2 
	\big)\epsilon_{1*}^2 \delta_{1*}
	+
	\big(
		\tfrac{37}{3}
		-
		\tfrac{130}{9}\, \ln 2  
		+
		3 \ln^2 2 
	\big)\epsilon_{1*} \delta_{1*}^2 
	\nonumber
	\\[0.5em]
	&\qquad
	-\,
	\big(
		\tfrac{308}{27}
		-
		\tfrac{5}{24}\, \pi^2  
		-
		16 \ln 2 
		+
		\tfrac{13}{2}\, \ln^2 2 
	\big)\epsilon_{1*} \epsilon_{2*} \delta_{1*}
	\nonumber
	\\[0.5em]
	&\qquad
	+\,
	\big(
		\tfrac{61}{18}
		-
		\tfrac{55}{6}\, \ln 2  
		+
		\tfrac{9}{2}\, \ln^2 2  
		-
		\tfrac{1}{2}\, \ln^3 2 
	\big)\epsilon_{2*} \delta_{1*}^2 
	-
	\big(
		\tfrac{13}{18}
		-
		\tfrac{11}{6}\, \ln 2  
		+
		\tfrac{5}{2}\, \ln^2 2  
		-
		\tfrac{1}{2}\, \ln^3 2 
	\big)\epsilon_{2*}^2 \delta_{1*} 
	\nonumber
	\\[0.5em]
	&\qquad
	-\,
	\big(
		\tfrac{14}{27}\,  
		-
		\tfrac{13}{72}\,\pi^2   
		-
		\tfrac{13}{9}\, \ln 2  
		+
		\tfrac{1}{24}\, \pi^2 \ln 2  
		+
		\tfrac{5}{2}\, \ln^2 2  
		-
		\tfrac{1}{2}\, \ln^3 2  
	\big)\epsilon_{2*} \epsilon_{3*} \delta_{1*}
	\nonumber
	\\[0.5em]
	&\qquad
	-\,
	\big(
		\tfrac{200}{27}
		-
		\tfrac{1}{6}\, \pi^2 
		-
		\tfrac{64}{9}\, \ln 2 
		+
		2 \ln^2 2 
	\big)\epsilon_{1*} \delta_{1*} \delta_{2*} 
	\nonumber
	\\[0.5em]
	&\qquad
	-\,
	\big(
		\tfrac{50}{27} 
		-
		\tfrac{1}{18}\, \pi^2  
		-
		\tfrac{43}{9}\,\ln 2  
		+
		\tfrac{1}{24}\, \pi^2 \ln 2  
		+
		3 \ln^2 2  
		-
		\tfrac{1}{2}\, \ln^3 2  
	\big)\epsilon_{2*} \delta_{1*} \delta_{2*} 
	\nonumber
	\\[0.5em]
	&\qquad
	-\,
	\big(
		\tfrac{361}{54}
		-
		\tfrac{15}{2}\, \ln 2   
		+
		\tfrac{13}{6}\,\ln^2 2   
		-
		\tfrac{1}{6}\, \ln^3 2  
	\big)\delta_{1*}^3 
	\nonumber
	\\[0.5em]
	&\qquad
	+\,
	\big(
		\tfrac{350}{27}
		-
		\tfrac{2}{9}\, \pi^2 
		-
		\tfrac{127}{9}\,\ln 2   
		+
		\tfrac{1}{24}\, \pi^2 \ln 2 
		+
		5 \ln^2 2  
		-
		\tfrac{1}{2}\, \ln^3 2 
	\big)\delta_{1*}^2 \delta_{2*} 
	\nonumber
	\\[0.5em]
	&\qquad
	-\,
	\big(
		3 
		-
		\tfrac{7}{72}\,\pi^2   
		-
		\tfrac{1}{4}\, \zeta(3)  
		-
		3  \ln 2 
		+
		\tfrac{1}{24}\, \pi^2 \ln 2  
		+
		\tfrac{7}{6}\,\ln^2 2   
		-
		\tfrac{1}{6}\, \ln^3 2  
	\big)\delta_{1*} \delta_{2*}^2
	\nonumber
	\\[0.5em]
	&\qquad
	-\,
	\big(
		3 
		-
		\tfrac{7}{72}\,\pi^2   
		-
		\tfrac{1}{4}\, \zeta(3)  
		-
		3 \ln 2  
		+
		\tfrac{1}{24}\, \pi^2 \ln 2  
		+
		\tfrac{7}{6}\, \ln^2 2  
		-
		\tfrac{1}{6}\, \ln^3 2  
	\big)\delta_{1*} \delta_{2*} \delta_{3*}
	\nonumber
	\\[0.5em]
	&\qquad
	-\,
	\tfrac{1}{9}
	\big(
		\epsilon_{2*} \epsilon_{3*}^2
		+
		\epsilon_{2*} \epsilon_{3*} \epsilon_{4*}
		+
		2\delta_{1*} \delta_{2*}^2 
		+
		2 \delta_{1*} \delta_{2*} \delta_{3*} 
	\big)\ln w.
\end{align}
As can be seen, the presence of terms involving $ \ln w $ in $ \bar r^{(3)} $ given above makes the fourth-order part of $ r $ logarithmically divergent.

\bigskip
\subsection{Spectral Index and Running of the Spectral Index}
\label{subsecSpecIndex}
The spectral index for scalar perturbation defined as
\begin{align}
	n_s - 1
	&\equiv
	\frac{\dd \ln P_s}{\dd \ln k},
\end{align}
can be computed from the derived expression for the power spectrum in Sec. \ref{logDiv} given by (\ref{powerSpec3}). The idea is that from the relation $k = -\nu_{s*}/(\eta_*c_{s*})$, telling us that $\eta_*$ is an implicit function of $k$, one can determine $\dd \ln (-\eta_{*})/\dd \ln k$ (see Refs. \cite{Martin:2013, Zhu:2014wfa}). The resulting expression can then be used to carry out the calculation of $n_s$ as dictated by the definition above. One may think that because the power spectrum under the current scheme involves logarithmic divergences starting at the third order, $n_s$ may inherit the same bad feature. However, $\eta$ and $k$ are independent in the limit as $\eta\rightarrow 0^-$ unless the turning point relating $k$ and $\eta_*$ happens to be at $\eta = \eta_* = 0$, which is a trivial scenario. Hence, even if we accept that the derived power spectrum containing logarithmically divergent parts is valid (which it is not), the resulting spectral index will still be finite.

Let us try to see the finiteness of the spectral indices under the current calculation scheme in a more rigorous way. We start with the definition of $n_s$ above with $P_s$ given by (\ref{powerSpec2}) and derive the appropriate working equation. The derivation of the working equation was done in Ref. \cite{Habib:2002yi} for the case $c_s^2 = 1$ and we are here going to follow the same path. Taking the logarithm of the power spectrum, we find
\begin{align}
	\frac{\dd \ln P_s}{\dd \ln k}
	&=
	3 + 2k\frac{\dd}{\dd k}\int _{\eta _*}^\eta 
	\dd\tau \,\sqrt{g_s(k,\tau )},
\end{align}
which implies upon using the definition of $ g_s $ given by (\ref{gScalar}),
\begin{align}
	n_s - 1
	&=
	3 - 2k^2\int _{\eta _*}^\eta \frac{\dd\tau\,c_s^2}{\sqrt{g_s(k,\tau )}}.
	\label{nsMinusOne}
\end{align}
This is our sought-for working equation for the spectral index. 

References \cite{Habib:2002yi,Habib:2004kc,Habib:2005mh} discussed the square-root singularity at the turning point in the integrand above for the case where $c_s^2 = 1$. For the case in hand, where $c_s^2 \ne \text{constant}$, we have the same form of singularity and we wish to find out using the expansion scheme used in Ref. \cite{Martin:2013} if the working equation above for $n_s$ leads to a finite or infinite results with logarithmic divergences just like that of the power spectrum. To proceed, we perform a Taylor series expansion in the same way we did in Sec. \ref{logDiv}.
\begin{align}
	\frac{c_s^2}{\sqrt{g_s(k,\tau )}}
	&=
	\frac{c_s^2}{\sqrt{g_s}}\bigg|_{\nu _{s*}, c_{s*}}
	+\,
	\frac{\partial }{\partial \nu _s}\frac{c_s^2}{\sqrt{g_s}}
	\bigg|_{\nu _{s*}, c_{s*}}\hspace{-1.5em}
	(\Delta \nu _s)
	\,+\,
	\frac{\partial }{\partial c_s^2}\frac{c_s^2}{\sqrt{g_s}}
	\bigg|_{\nu _{s*}, c_{s*}}\hspace{-1.5em}
	(\Delta c_s^2)
	+
	\cdots.
	\label{taylorSOverG}
\end{align}
The first term in the right hand side corresponds to a convergent integral. The rest of the terms can be classified into three types: those involving (a) $(\Delta c_s^2)^n$ only, (b) $(\Delta \nu_s)^n$ only, and (c) $(\Delta \nu_s)^n(\Delta c_s^2)^p$, where $n,\,p \ge 1$.

For terms involving \textit{only} $ (\Delta c_s^2)^n $, with $ n\ge 1 $, the corresponding integral takes the form
\begin{align}
	\int _1^w
	\dd u 
	\left[	
	\frac{c_1\,u^{2n-1}\ln^m u}{\big(1-u^2\big)^{n-\frac{1}{2}}}
	+	
	\frac{c_2\,u^{2n+1}\ln^m u}{\big(1-u^2\big)^{n+\frac{1}{2}}}
	\right],
	\qquad
	(m \ge n)
\end{align}
where $ c_1 $ and $ c_2 $ are constants. The first term inside the pair of square brackets is continuous on $ (0,1) $. Furthermore, the limits as $ u\rightarrow 0^+ $ and $ u\rightarrow 1^- $ both exist and vanishing. The same goes for the second term except, when $ m = n $, where the limit as $ u\rightarrow 1^- $ diverges. This however, does not lead to a divergent integral. That this holds can be seen by using a comparison test for improper integrals with the case for $n = 1$ as the basis. The second term can result to a divergent integral if $ m < n $ which is clearly not the case here. 

For terms involving \textit{only} $ (\Delta \nu _s)^n $, with $ n \ge 1 $, the corresponding integral takes the form 
\begin{align}
	\int _1^w\dd u
	\left[
		\frac{c_1u\ln^m u}{\big(1 - u^2\big)^{n + \frac{1}{2}}}
		+
		\frac{c_2u\ln^m u}{\big(1 - u^2\big)^{n - \frac{1}{2}}}
		+
		\cdots
	\right],
	\qquad
	(m \ge n)
\end{align}
The number of terms inside the pair of square brackets depends on whether $ n $ is odd or even. If it is odd the number of terms is $ (n + 1)/2 $. If it is even the number of terms is $ (n + 2)/2 $. All the terms inside the pair of square brackets are continuous on $ (0, 1) $ and the limits as $ u \rightarrow 0^+ $ and $ u\rightarrow1^- $ both exist and vanishing except for the first one where the limit as $ u\rightarrow 1^- $ diverges. However, as in the case of terms involving $ (\Delta c_s^2)^n $, this does not lead to a divergent integral. That this holds can be seen by using a comparison test for improper integrals with the case for $n = 1$ as the basis. The integral can diverge if $ m < n $ which is clearly not the case here.

The analysis for cross terms in the Taylor series expansion given by (\ref{taylorSOverG}) involving $ (\Delta \nu _s)^n(\Delta c_s^2)^p $, with $ n,\,p \ge 1 $, leads to the same finding that the corresponding integral is convergent. This together with the results above for the cases of $(\Delta c_s^2)^n$ and $(\Delta \nu_s)^n$, leads us to conclude that the special form of the working equation for $ n_s $ given by (\ref{nsMinusOne}) enables the calculation the spectral index to any order with respect to the Hubble and sound flow parameters without encountering logarithmic divergences involving the limit as $w\rightarrow 0^+$ of $\ln w$.

The examination for the spectral index for tensor perturbation follows the same route. From its definition given by
\begin{align}
    n_t - 1
    &\equiv
    \frac{\dd \ln P_t}{\dd \ln k},
    \label{tensorNt}
\end{align}
we arrive at the working equation similar to that of (\ref{nsMinusOne}) for the scalar perturbation.
\begin{align}
	\label{nstMinusOne}
    n_t - 1
    &=
    3
    -
    2k^2\int_{\eta_*}^\eta 
    \frac{\dd \tau}{\sqrt{g_t(k,\tau)}}.
\end{align}
As in the case of scalar perturbations, the placement of $\sqrt{g_t}$ being the denominator of the integrand makes the integral above convergent. More formally, the Taylor series expansion of the integrand takes the same form as that of the scalar perturbation with $c_s^2 = 1$ because the index functions $\nu_s$ and $\nu_t$ have the same form. With $n_s$ being computable under the current scheme up to any order with respect to $(\epsilon_{n*},\, \delta_{n*})$, the same holds for $n_t$.

The fact that $n_s$ and $n_t$ can be computed to any order tells us that the running of these spectral indices defined as
\begin{align}
	\alpha _s
	&\equiv
	\frac{\dd n_s}{\dd \ln k},
	\qquad
	\alpha _t
	\equiv
	\frac{\dd n_t}{\dd \ln k},
\end{align}
for scalar and tensor perturbations respectively, can as well be calculated to any order in $(\epsilon_{n*},\, \delta_{n*})$. One method, already elaborated at the beginning of this subsection, is to simply use the relation $k = -\nu_{s*}/(\eta_*c_{s*})$ for scalar perturbation to find $\dd\ln(-\eta_*)/\dd\ln k$ and then substitute the result together with the expression for $n_s$ coming from (\ref{nsMinusOne}) in the definition given above for $\alpha_s$. An equivalent procedure holds for the tensor perturbation. The other method involves directly differentiating the expression for $n_s$ given by (\ref{nsMinusOne}) and deriving the working equation for $\alpha_s$ in the same way we did for $n_s$. However, the resulting expression for this one involves an integrand with $g_s^{\frac{3}{2}}$ being part of the denominator different from that of $g_s^{\frac{1}{2}}$ in the expression for $n_s$. This leads to some practical difficulties because of the resulting divergent integrals.

\bigskip
\bigskip
\section{Remarks on Logarithmic Divergences}
\label{logDivRamification}
In what follows we clarify some of the questions or reiterate some of the main points stemming from the emergence of logarithmic divergences that we point out in this work.

\textit{The finite lower-order results}. Given the exact same method to calculate the $ \text{first-} $, second-, and so on, order part of a given physical quantity, (in fact, they are all tied together,) the emergence of logarithmic divergences say, in the third-order part, questions the validity of the lower-order parts even though they converge at second order. As such, the ``consistency check'' in the calculation of the power spectra up to \textit{second order} carried out in Ref. \cite{Martin:2013}, for which all terms that can lead to logarithmic divergence canceled, should be viewed with caution; as we have demonstrated, all terms starting at the third order do not pass this ``consistency check.'' It is in this same spirit that the results of calculations carried out in Refs. \cite{Lorenz:2008et} and \cite{Zhu:2014wfa} should be taken with caution. Moreover, it may be worth pointing out that even as far as the third order of precision in the UA, the percent difference of some of the coefficient of the second-order combinations of Hubble flow parameters, with respect to that obtained using other approximation methods, can still be significant; e.g., for $ \epsilon _{1\star}\epsilon _{2\star} $ in the expression for the scalar and tensor power spectra, where the symbol $\star$ means evaluation at the horizon crossing, the percent difference with respect to that obtained through the Green's function method is 90\% and 13\%, respectively (see Table III of Ref. \cite{Zhu:2014wfa}). The resulting practical expressions for the power spectra and tensor-to-scalar ratio do not rest on a solid mathematical foundation and hence, are in need of some further justification.

We note that this work can only go this far. There is still a substantial possibility that the inconsistency in the expansion scheme can only affect some but not all of the finite first- and second-order terms for the power spectra and the finite first- to third-order terms in the tensor-to-scalar ratio. A thorough check should be put in place. In order to do this, one may need to carry out the calculations well beyond the third order of precision in the UA. The succession of the computations however, for the power spectra, from Ref. \cite{Lorenz:2008et} (first order in Hubble and sound flow parameters and first order in the UA) and Ref. \cite{Martin:2013} (second order in Hubble and sound flow parameters and first order in the UA), to Ref. \cite{Zhu:2014wfa} (second order in Hubble and sound flow parameters and third order in the UA), tells us that this is not an easy task. The enormity of the calculations involved, the algorithms to be designed to make it practically doable, makes such a task too complex to be included here; it deserves a separate study of its own\footnote{There is a way (although also tedious,) to possibly disprove the lower order finite results for the power spectra : Compute $\dd \ln P_s/\dd \ln k$ (or $\dd \ln P_t/\dd \ln k$) and compare the result with that calculated through the working equation for the corresponding spectral index. Note that although it can disprove the former if the two results do not match, it might not be able to prove the validity of the lower-order finite results for the power spectra.}.

\textit{The all-order finite quantities}. We deal in this study with seven quantities namely, power spectra (2), tensor-to-scalar ratio (1), spectral indices (2), and running of the spectral indices (2). The first three exhibit logarithmic divergences while the last four do not. The working equation for the power spectra shares the same form. It follows that, should the logarithmic divergence appear in $ P_s $, we expect it to also appear in $ P_t $ unless the divergence is only tied to the speed of sound which is clearly not the case here. The tensor-to-scalar ratio being the  quotient of two divergent quantities, is likely to inherit such divergence; this, we find to be the case. Now, as can be traced in Subsec. \ref{subsecSpecIndex}, in comparison to that of Subsec. \ref{uniformApprox} for the power spectra, the remaining four quantities in our list share a different structure of the working equation making them compatible with the expansion scheme. In particular, in accord with the calculation in Sec. \ref{subsecSpecIndex}, with the placement of $ \sqrt{g_s} $ ($ \sqrt{g_t} $) as the denominator of the integrand for $ n_s $ ($ n_t $), the end result is finite. Thus, these all-order finite quantities (spectral indices and running of the spectral indices), apart from the possible imprecision associated with the preliminary exclusion of some correction factors in the UA, are valid.

\textit{Origin of the divergences}. The logarithmic divergences come from the incomplete cancellation of logarithmic terms brought about by the expansion scheme (laid down in \cite{Martin:2013} and elaborated in Subsection \ref{seriesExpandOther}) used in the calculation of some physical quantities such as the power spectra and tensor-to-scalar ratio, through the UA. Such divergences cannot be attributed to the prescription involving taking the limit\footnote{More technically, it is $k\eta\,\rightarrow\, 0^-$. However, $k$ cancels out for every pair $(k,\eta)$ or $(k,\tau)$; for instance, in (\ref{uws}), $\ln w = \ln \frac{k\eta}{k\eta_*} = \ln \frac{\eta}{\eta_*}.$} as $ \eta\,\rightarrow\, 0^- $ in the working equations (see for example, (\ref{scalarPii}) and (\ref{tensorP})) for the power spectra among others, because these working equations are mathematically well-defined in the mentioned limit. As such, the divergences are not a consequence of the UA itself but a mathematical artifact of the expansion scheme used in the 
UA for the computation of the power spectra and tensor-to-scalar ratio. 

This implies that the expansion scheme has a characteristic that makes it incompatible with the working equation for the power spectra (and for the tensor-to-scalar ratio being just that---the ratio of the two power spectra). Such a characteristic cannot be attributed to the way we expand based on the variable of differentiation. Consider for instance, (\ref{eps1Expand}) for the first slow-roll parameter.
\begin{align*}
\epsilon_1
=
\epsilon_{1*} - \epsilon_{1*}\epsilon_{2*}\ln\frac{\eta}{\eta_*} + \cdots
\end{align*}
This end result is the same whether we use (\ref{epsilonExpand}) involving $N$ or the more computationally cumbersome (see Sec. \ref{logDiv}) expansion exemplified by (\ref{altExpand}) involving $\eta$:
\begin{align*}
\epsilon_1
&=
\epsilon_{1*}
+
\frac{\dd\epsilon_1}{\dd N}\bigg|_*(\Delta N)
+
\frac{1}{2!}
\frac{\dd^2\epsilon_1}{\dd N^2}\bigg|_*(\Delta N)^2
+
\cdots
\\[0.5em]
\epsilon_1
&=
\epsilon_{1*}
+
\frac{\dd\epsilon_1}{\dd \eta}\bigg|_*(\eta - \eta_*)
+
\frac{1}{2!}
\frac{\dd^2\epsilon_1}{\dd N^2}\bigg|_*(\eta - \eta_*)^2
+
\cdots,
\end{align*}
However, note that many (in fact, infinite) combinations  of the Hubble and sound flow parameters are of the same order; e.g., $(\epsilon_{1*}^2, \epsilon_{1*}\epsilon_{2*}, \epsilon_{2*}\epsilon_{3*},\cdots) \sim \mathcal O(\epsilon^2)$. This gives us a hint that although the expansion for a given background quantity with respect to the Hubble and sound flow parameters does not depend on the variable of differentiation in the corresponding Taylor series, the expansion itself may not be unique. And if it is not unique and at the same time incompatible with the working equation for the power spectra, then the Hubble and sound flow parameters may not be the appropriate perturbation expansion parameter for the power spectra ($P_s$ and $P_t$). The characteristic we are looking for, then, is the base of the expansion itself---the Hubble and sound flow parameters.

We recall that unlike that of a typical perturbed equation of motion (e.g., that of a simple harmonic oscillator whose potential has an added small quartic term involving displacement of the oscillator about a certain equilibrium position), the Mukhanov-Sasaki equation does not contain an explicit perturbation parameter. As such, one should be careful about using any expansion parameter for any quantity whether it be physical or not, defined or derived from the Mukhanov-Sasaki equation. The suitability of an expansion scheme with respect to a given set of expansion parameters rests on the structure of a given (approximating) equation for a given quantity (in relation to the Mukhanov-Sasaki equation). For the case in hand, as already pointed out above, the current expansion scheme is incompatible with the working equations for the power spectra and compatible with that of the spectral indices. In relation to this, we note that there may be another group of perturbation expansion parameters with respect to which the power spectra is finite to all orders. Whether such finite results would agree and to what extent, with the finite lower-order results for the power spectra given in Sec. \ref{logDiv}, remains to be seen.

\textit{Physicality of the divergences}. There are a multitude of ways by which a physical interpretation can be attached to the logarithmic divergences involved in this work. The expansion for the first slow-roll parameter for instance, $\epsilon_1 = \epsilon_{1*} - \epsilon_{1*}\epsilon_{2*}\ln(\eta/\eta_*) + \cdots,$ may be seen as an improper characterisation of the background. That this does not work for the power spectrum and works for the spectral index, seems to contradict this interpretation. We take the view that they are primarily mathematical in nature. Prematurely attaching a physical meaning may only complicate the process of solving the problem (in future work) brought about by the infinite final results. It is the mathematical aspect of the implementation of the UA that exhibits the problem and not the physics that it is trying to describe.

\textit{Resolution in sight}. Since the origin is mainly of mathematical nature namely, an artifact of an expansion scheme, the most probable resolution in sight should arise with respect to the same nature---find another expansion scheme. Such an expansion scheme should avoid as much as is (computationally) practicable, terms\footnote{Needless to say, we are not closing the possibility that a similar expansion scheme involving terms such as $ \ln (\eta/\eta_*) $ may be able to lead to a finite result. In such a case, a complete cancellation of all potentially divergent terms is all that is needed.} that can potentially lead to a divergent final result. Moreover, it should retain the pragmatic decomposition of the background quantities (e.g., slow-roll parameter, scale factor, speed of sound, etc) in ascending order with respect to some set of parameters; see for instance, the expansion of $ \epsilon _1 $ given by (\ref{eps1Expand}). This should be contrasted with the (impractical) alternative expansion scheme exemplified by (\ref{altExpand}) where the higher-order derivative terms involve a mix of lower- and higher-order sub-terms in the Hubble and sound flow parameters.

\bigskip
\bigskip
\section{Concluding Remarks}
\label{conc}
The uniform approximation (UA) is one of the leading semi-analytical approaches used to solve the Mukhanov-Sasaki equation and hence, to calculate relevant physical quantities in inflationary cosmology. It is a systematically improvable scheme with finite error bounds and does not require constant slow-roll parameters \cite{Habib:2002yi}. Nonetheless, the practical implementation of this method at least for slow-roll \textit{k}-inflation, relies on a suitable expansion scheme in order to express physical quantities in terms of some possibly measurable parameters.

We note the expansion scheme for slow-roll parameters and other background quantities (e.g., scale factor, speed of sound, etc) introduced in \cite{Lorenz:2008et}, expounded in Ref. \cite{Martin:2013}, and adopted in Refs. \cite{Zhu:2014aea,Zhu:2014wfa}. This expansion scheme involves powers of $ \ln(\eta/\eta_*) $. With the prescription that the limit as $ \eta\rightarrow 0^- $ has to be evaluated at the end of the calculation, this may potentially lead to logarithmically divergent result. In their calculation of the power spectra up to second order with respect to the Hubble and sound flow parameters, the authors of Ref. \cite{Martin:2013} stated a ``consistency check'' as to the validity of their implementation of the UA that can be stated as : all terms that can potentially give rise to logarithmically divergent result should exactly cancel prior to taking the limit as $ \eta\rightarrow 0^- $. Their second-order calculation successfully passed this ``consistency check,'' but the all-order check was not carried out. There was also no rigorously sufficient characterisation of all forms of divergences that may come out of the calculation. (This seems unnecessary as long as there is an assurance that all forms of divergences will cancel out in the end; it turns out not to be the case.) 

This work is an attempt to fill in these gaps. Focusing on slow-roll $k$-inflation, we (a) characterised the form of the divergences as \textit{purely} logarithmic in nature and (b) set the limit as to how far one can calculate several physical quantities in terms of the Hubble and sound flow parameters before hitting a logarithmically divergent result. It turns out that the power spectra for both the scalar and tensor perturbations can only be calculated up to second order in terms of the Hubble and sound flow parameters (see Table \ref{logDivTable}). The tensor-to-scalar ratio is likewise bounded up to third order. However, the spectral indices and the running of the spectral indices can be calculated to any order. So far, the most precise calculations of these physical quantities in $k$-inflation through the UA can be found in Ref. \cite{Zhu:2014wfa} where the calculation is carried out up to third order of precision in the UA. The boundary that we set here is consistent with this work.

\begin{table}[h]
\bigskip
\centering
\begin{tabular}{llllll}
\textbf{Physical Quantity}     & 0 & 1 & 2 & 3 & 4 \\[0.5em] \hline \\[-0.50em]
power spectrum (scalar), $P_s$      & $\circ$ & $\circ$  & $\circ$  & $\bullet$  & $\bullet$  \\ 
power spectrum (tensor), $P_t$      & $\circ$  & $\circ$  & $\circ$  & $\bullet$  & $\bullet$  \\ 
tensor-to-scalar ratio, $r$         &   & $\circ$  & $\circ$  & $\circ$  & $\bullet$  \\ 
spectral index (scalar), $n_s$      & $\circ$  & $\circ$  & $\circ$  & $\circ$  & $\circ$  \\ 
spectral index (tensor), $n_t$      & $\circ$  & $\circ$  & $\circ$  & $\circ$  & $\circ$  \\ 
running (scalar), $\alpha_s $ & $\circ$  & $\circ$  & $\circ$  & $\circ$  & $\circ$  \\ 
running (tensor), $\alpha_t$ & $\circ$  & $\circ$  & $\circ$  & $\circ$  & $\circ$ \\ 
\end{tabular}
\caption{Logarithmic divergences in the calculation of the power spectrum and other physical quantities. Filled circles represent results involving logarithmically divergent terms while unfilled circles represent finite results. The numbers in the first row indicate the order of a given physical quantity with respect to the Hubble and sound flow parameters. Note that once a logarithmic divergence appears (as in the power spectra and tensor-to-scalar ratio), it permeates to all succeeding orders. The spectral indices and running are finite to all orders.}
\label{logDivTable}
\bigskip
\end{table}

Looking back, we find nothing inherently wrong with the mathematically well-defined working equations provided by the UA for the power spectra and tensor-to-scalar ratio. It is with the practical implementation of these working equations that we get the logarithmically divergent results in the higher-order parts.  Such divergences stem from the use of the expansion scheme involving potentially divergent terms that do not find a complete cancellation in the end. It turns out that with respect to the structure of the working equation for the seven quantities listed in Table \ref{logDivTable}, the expansion scheme is incompatible with that of the power spectra and tensor-to-scalar ratio and compatible with that of the spectral indices and running of these indices. The calculation results for the last four quantities ($ n_s,\, n_t,\, \alpha _s,\, \alpha _t $) are therefore, completely valid apart from possible imprecision that can be remedied by increasing the level of precision of calculation in the UA. The first three quantities on the other hand, do not rest on a solid mathematical foundation. For this reason, some of the terms in the lower-order finite results for $ P_s $, $ P_t $, and $ r $ may be misleading. However, owing to the enormity of the calculations involved needed to verify this, we set aside the necessary analysis of these lower order terms for future studies.

We hope that we have clarified the limits of one significant implementation of the UA in the calculation of several important physical quantities in inflationary cosmology. Ours is a rigorous identification and characterisation of the problem in relation to the existence of logarithmic divergences in an otherwise, finite set of quantities. In our view this puts the current works in the literature into a proper perspective. Moreover, this sets up the framework of how to solve a problem that has only just been formally identified and characterised in this work: How do we implement the UA in the calculation of power spectra for slow-roll \textit{k}-inflation without suffering from logarithmic divergences to all orders with respect to a given set of perturbation expansion parameters?

\bigskip
\bigskip
\acknowledgments

We would like to thank members of the Particle Physics Theory Group (Osaka University); in particular, Yukari Nakanishi for useful discussions on the UA and Carsten Fritzner Fr{\o}strup (also of the University of Copenhagen) for assistance in the preparation of the manuscript.

\bigskip
\bigskip


\begin{thebibliography}{99}

\bibitem{Starobinsky:1979ty}
  A.~A.~Starobinsky,
  JETP Lett.\  {\bf 30} (1979) 682
   [Pisma Zh.\ Eksp.\ Teor.\ Fiz.\  {\bf 30} (1979) 719].
  
\bibitem{Guth:1980zm}
  A.~H.~Guth,
  Phys.\ Rev.\ D {\bf 23} (1981) 347.
  
\bibitem{Linde:1981mu}
  A.~D.~Linde,
  Phys.\ Lett.\ B {\bf 108} (1982) 389.

\bibitem{Albrecht:1982wi}
  A.~Albrecht and P.~J.~Steinhardt,
  Phys.\ Rev.\ Lett.\  {\bf 48} (1982) 1220.
  
\bibitem{LiddlenLyth}
  A.~R.~Liddle, D.~H.~Lyth, 
  ``The primordial density perturbation: Cosmology, inflation and the origin of structure,'' 
  NY, USA: Cambridge Univ. Pr. (2009) 516 p.
  
\bibitem{Mukhanov:2005sc}
  V.~Mukhanov,
  ``Physical foundations of cosmology,''
  Cambridge, UK: Univ. Pr. (2005) 421 p
  
\bibitem{Martin:2014}
  J.~Martin, C.~Ringeval and V.~Vennin,
  Phys.\ Dark Univ.\  (2014)
  [arXiv:1303.3787 [astro-ph.CO]].
  
\bibitem{Albrecht:2002uz}
  A.~Albrecht,
  In *Barrow, J.D. (ed.) et al.: Science and ultimate reality* 363-401
  [astro-ph/0210527].

\bibitem{Parker:2007}
  L.~Parker,
  hep-th/0702216 [HEP-TH].
    
\bibitem{Bastero-Gil:2013}
  M.~Bastero-Gil, A.~Berera, N.~Mahajan and R.~Rangarajan,
  Phys.\ Rev.\ D {\bf 87} (2013) 8,  087302
  [arXiv:1302.2995 [astro-ph.CO]].
  
\bibitem{Alinea:2015pza}
  A.~L.~Alinea, T.~Kubota, Y.~Nakanishi and W.~Naylor,
  JCAP {\bf 06} (2015) 019
  [arXiv:1503.08073 [gr-qc]].

\bibitem{Bartolo:2004if}
  N.~Bartolo, E.~Komatsu, S.~Matarrese and A.~Riotto,
  Phys.\ Rept.\  {\bf 402} (2004) 103
  [astro-ph/0406398].

\bibitem{Martin:2013}
  J.~Martin, C.~Ringeval and V.~Vennin,
  JCAP {\bf 1306} (2013) 021
  [arXiv:1303.2120 [astro-ph.CO]].
  
\bibitem{Zhu:2014aea}
  T.~Zhu, A.~Wang, G.~Cleaver, K.~Kirsten and Q.~Sheng,
  Phys.\ Rev.\ D {\bf 90} (2014) 6,  063503
  [arXiv:1405.5301 [astro-ph.CO]].
  
\bibitem{Lorenz:2008et}
  L.~Lorenz, J.~Martin and C.~Ringeval,
  Phys.\ Rev.\ D {\bf 78} (2008) 083513
  [arXiv:0807.3037 [astro-ph]].
  
\bibitem{Zhu:2014wfa} 
  T.~Zhu, A.~Wang, G.~Cleaver, K.~Kirsten and Q.~Sheng,
  Phys.\ Rev.\ D {\bf 90}, no. 10, (2014) 103517
  [arXiv:1407.8011 [astro-ph.CO]].
  
\bibitem{Miller:2006}
  P.~T.~Miller, 
  American Mathematical Society, USA: (2006) 467p.
  
\bibitem{Olver}
  F.~W.~J.~Olver, 
  AK Peters, USA: (1997) 547p.
    
\bibitem{ArmendarizPicon:1999rj}
  C.~Armendariz-Picon, T.~Damour and V.~F.~Mukhanov,
  Phys.\ Lett.\  B {\bf 458} (1999) 209
  [arXiv:hep-th/9904075].

\bibitem{Garriga:1999vw}
  J.~Garriga and V.~F.~Mukhanov,
  Phys.\ Lett.\  B {\bf 458} (1999) 219
  [arXiv:hep-th/9904176].
  
\bibitem{Bardeen:1983}
  J.~M.~Bardeen, P.~J.~Steinhardt, and M.~S.~Turner,
  Phys.\ Rev.\ D{\bf 28} (1983) 679.
  
\bibitem{Habib:2002yi}
  S.~Habib, K.~Heitmann, G.~Jungman and C.~Molina-Paris,
  Phys.\ Rev.\ Lett.\  {\bf 89} (2002) 281301
  [astro-ph/0208443].
  
\bibitem{Habib:2004kc}
  S.~Habib, A.~Heinen, K.~Heitmann, G.~Jungman and C.~Molina-Paris,
  Phys.\ Rev.\ D {\bf 70} (2004) 083507
  [astro-ph/0406134].
  
\bibitem{Zhu:2013upa}
  T.~Zhu, A.~Wang, G.~Cleaver, K.~Kirsten and Q.~Sheng,
  Phys.\ Rev.\ D {\bf 89} (2014) 4,  043507
  [arXiv:1308.5708 [astro-ph.CO]].
  
\bibitem{Rojas:2011rg}
  C.~Rojas and V.~M.~Villalba,
  JCAP {\bf 1201} (2012) 003
  [arXiv:1112.1342 [astro-ph.CO]].
  
\bibitem{Arnowitt:1959ah} 
  R.~L.~Arnowitt, S.~Deser and C.~W.~Misner,
  Phys.\ Rev.\  {\bf 116} (1959) 1322.
  
\bibitem{Chen:2010xka} 
  X.~Chen,
  Adv.\ Astron.\  {\bf 2010} (2010) 638979
  [arXiv:1002.1416 [astro-ph.CO]].

\bibitem{Bruneton:2006gf} 
  J.~P.~Bruneton,
  Phys.\ Rev.\ D {\bf 75} (2007) 085013
  [gr-qc/0607055].
  
\bibitem{Lucchin:1984yf}
  F.~Lucchin and S.~Matarrese,
  Phys.\ Rev.\ D {\bf 32} (1985) 1316.
  
\bibitem{Martin:2000ei} 
  J.~Martin and D.~J.~Schwarz,
  Phys.\ Lett.\ B {\bf 500} (2001) 1 
  [astro-ph/0005542].

\bibitem{Stewart:1993bc} 
  E.~D.~Stewart and D.~H.~Lyth,
  Phys.\ Lett.\ B {\bf 302} (1993) 171 
  [gr-qc/9302019].
  
\bibitem{Gong:2001he} 
  J.~O.~Gong and E.~D.~Stewart,
  Phys.\ Lett.\ B {\bf 510} (2001) 1 
  [astro-ph/0101225].

\bibitem{Lidsey:1995np} 
  J.~E.~Lidsey, A.~R.~Liddle, E.~W.~Kolb, E.~J.~Copeland, T.~Barreiro and M.~Abney,
  Rev.\ Mod.\ Phys.\  {\bf 69} (1997) 373 
  [astro-ph/9508078].
    
\bibitem{Martin:1999wa} 
  J.~Martin and D.~J.~Schwarz,
  Phys.\ Rev.\ D {\bf 62} (2000) 103520 
  [astro-ph/9911225].
  
\bibitem{Schwarz:2001vv} 
  D.~J.~Schwarz, C.~A.~Terrero-Escalante and A.~A.~Garcia,
  Phys.\ Lett.\ B {\bf 517} (2001) 243
  [astro-ph/0106020].
  
\bibitem{Gong:2004kd}
  J.~O.~Gong,
  Class.\ Quant.\ Grav.\  {\bf 21} (2004) 5555
  [gr-qc/0408039].
  
\bibitem{Martin:2002vn}
  J.~Martin and D.~J.~Schwarz,
  Phys.\ Rev.\ D {\bf 67} (2003) 083512
  [astro-ph/0210090].
  
\bibitem{Casadio:2005xv}
  R.~Casadio, F.~Finelli, M.~Luzzi and G.~Venturi,
  Phys.\ Lett.\ B {\bf 625} (2005) 1
  [gr-qc/0506043].
  
\bibitem{Casadio:2005em}
  R.~Casadio, F.~Finelli, M.~Luzzi and G.~Venturi,
  Phys.\ Rev.\ D {\bf 72} (2005) 103516
  [gr-qc/0510103].
  
\bibitem{Lebedev:1996}
  V.~I.~Lebedev,
  ``An introduction to functional analysis in computational mathematics,''
  {USA: Birkhäuser (1996) 256p}
  
\bibitem{Habib:2005mh}
  S.~Habib, A.~Heinen, K.~Heitmann and G.~Jungman,
  Phys.\ Rev.\ D {\bf 71} (2005) 043518
  [astro-ph/0501130].

\end{thebibliography}
\end{document}